%% file: paper_v8.tex
\documentclass[journal]{IEEEtran}
\usepackage{color}     
\usepackage{epsf,psfrag,amssymb,amsfonts,latexsym,cite,verbatim,enumerate,subcaption}
\usepackage{srcltx,amsmath,cases,graphicx}
\usepackage{times, multirow, array}
\usepackage[mathscr]{eucal}
\usepackage{cancel}
 \usepackage{algorithm}
\usepackage{algpseudocode}

\input macros

\input labelfig.tex

\newcommand {\Ebb}{{\mathbb{E}}}

\newtheorem{problem}{Problem}
\newtheorem{proposition}{Proposition}

\algnewcommand\algorithmicinput{\textbf{Step}}
\algnewcommand\Step{\item[\algorithmicinput]}

\begin{document}

\title{Tomlinson-Harashima Precoding-Aided Multi-Antenna Non-Orthogonal Multiple Access}

\author{
Jungho So, {\em Student~Member, IEEE}, Youngchul Sung$^\dagger$\thanks{$^\dagger$Corresponding author}, {\em Senior~Member, IEEE},\\ and Yong H. Lee, {\em Senior~Member, IEEE} \\
\thanks{The authors are with School of Electrical Engineering,  KAIST, Daejeon 305-701, South Korea. E-mail: jhso@kaist.ac.kr;  ysung@ee.kaist.ac.kr; yohlee@kaist.ac.kr.  This work was supported in part by 'The Cross-Ministry Giga KOREA Project' grant funded by the Korea government (MSIT) (No.GK17N0100, 5G Mobile Communication System Development based on mmWave) and in part  by 'The Cross-Ministry Giga KOREA Project' grant funded by the Korea government (MSIT) (No.GK17S0400, Research and Development of Open 5G Reference Model).}
}

\markboth{\protect\footnotesize Submitted to {\it IEEE JOURNAL OF SELECTED TOPICS IN SIGNAL PROCESSING}, \today}{So,  Sung and Lee}

\maketitle

\begin{abstract}
In this paper, Tomlinson-Harashima Precoding (THP) is considered for  multi-user multiple-input single-output (MU-MISO) non-orthogonal multiple access (NOMA) donwlink. Under the hierarchical structure in which multiple clusters each with two users are formed and served in the spatial domain and users in each cluster are served in the power domain, THP is applied to eliminate the inter-cluster interference (ICI) to the strong users and enlarge the dimension of the beam design space for mitigation of ICI to weak users as compared to conventional zero-forcing (ZF) inter-cluster beamforming.
With the enlarged beam design space, two beam design algorithms for THP-aided MISO-NOMA  are proposed. The first is a greedy sequential beam design with user scheduling, and the second is the joint beam redesign and power allocation. The two design problems lead to non-convex optimization problems. An efficient algorithm is proposed to solve the non-convex optimization problems based on successive convex approximation (SCA).
Numerical results show that the proposed user scheduling and two beam design methods based on THP yield noticeable gain over existing methods.
\end{abstract}


%
\IEEEpeerreviewmaketitle

\section{Introduction}
\label{sec:intro}

Power-domain NOMA also known as multi-user superposition transmission (MUST) is one of the promising technologies for 5G wireless communication to enhance the spectral efficiency \cite{3gpp:NOMA,Saito:13VTC}. Conventionally, the wireless communication resources  such as time, bandwidth and spatial domains were divided into multiple orthogonal resource blocks, and one user is assigned to each orthogonal resource block. Unlike such conventional orthogonal multiple access, in NOMA the base station (BS) serves multiple users in a single orthogonal resource block  based on superposition coding and successive interference cancellation (SIC) by exploiting the power domain.  Initially, NOMA was studied for single-input single-output (SISO) systems \cite{Ding:14SPL,Timotheou:15SPL, Liu15:PIMRC,So&Sung:15ComLet}, but recently there have been extensive research works to extend NOMA to multiple-antenna systems \cite{Kim13:MILCOM,Lan14:ICSPCS,Liu15:ICCW,Hanif16:TSP,Ding&Adachi&Poor:16WC,Sayed17:WD,Seo&Sung:18SP}. NOMA in multiple-antenna systems is attractive since it  increases the spectral efficiency further on top of the multiple antenna technology. On the contrary to single-antenna NOMA in which only the power domain exists,  in multiple-antennna NOMA there exist spatial and power domains to be exploited for user multiplexing and data transmission. These joint spatial and power domains should be used efficiently for  operation of multiple-antenna NOMA, and corresponding user scheduling/grouping,  beam design and power allocation are of great importance for good performance of multiple-antenna NOMA.

\subsection{Related Works and Motivation}
\label{subsec:RelatedWork}

In this paper, we  consider the MU-MISO NOMA downlink. Although there exists vast literature for MU-MISO NOMA, we discuss only the most related works to our work   in this subsection.
 Even though the problem of supporting multiple users in MU-MISO NOMA downlink can be approached by multi-user  SIC without hierarchy as in \cite{Hanif16:TSP}, we here consider the hierarchical approach as in
 \cite{Kim13:MILCOM,Liu15:ICCW,Ding&Adachi&Poor:16WC,Sayed17:WD,Seo&Sung:18SP}, which  is simple and attractive from the perspective of design and SIC complexity.
In  the hierarchical approach,   simultaneously-served users  are first grouped into multiple clusters, and then multiple clusters are supported in the spatial domain  while users  in each cluster are supported in the power domain. That is, users in each cluster  share the same spatial beam vector and are supported by superposition coding and SIC. In these works as well as in many other MU-MISO NOMA works, researchers assume that two users (one strong user and one weak user) are grouped in each cluster, considering signaling overhead and SIC error propagation \cite{Kim13:MILCOM,Liu15:ICCW,Chen16:Access,Chen16:TSP,Sayed17:WD,Seo&Sung:18SP}. We make this assumption too in this paper.  Then, the main problem in the hierarchical approach to MU-MISO NOMA is the joint design of beams, power allocation and user scheduling. However, this joint design of beams, power allocation and  user scheduling  is a complicated problem.
 Hence,
under the assumption of two-user grouping for each cluster, to make the joint design problem tractable, the step of beam design for MU-MISO NOMA was simplified by designing the beams as  linear ZF beams based on strong users' channels, and then  two users in each cluster share  the same beam \cite{Liu15:ICCW,Sayed17:WD}.  The reason for this beam design strategy is to keep up to the asymmetric NOMA principle that the strong users having high-quality channels are not limited by noise or interference, whereas the weak users having bad channels are limited by noise.
 Under this hierarchical MU-MISO NOMA structure  with the beams designed as ZF beams for strong users, several user scheduling/grouping and/or power allocation methods were proposed \cite{Kim13:MILCOM,Liu15:ICCW,Sayed17:WD,Seo&Sung:18SP}.
Although this linear ZF beam design strategy simplifies the overall problem for the hierarchy-based MU-MISO NOMA, it has limitation.
Suppose that we have $N_t$ transmit antennas and $N_t$ clusters each consisting of one strong user and one weak user. In case that the beams are designed as the ZF beams based on the strong users' channels, the beam for each cluster should be in the one-dimensional orthogonal space of the linear space spanned by the remaining $N_t-1$ clusters' strong users' channels. Hence, there is no freedom in the beam design and ICI is controlled solely by weak user selection. Thus, by enlarging the dimension of the beam design space, we can improve the weak user performance in addition to weak user selection.

\subsection{Contributions of the Paper}
\label{subsec:cont}

 In this paper, we consider the aforementioned hierarchical design fo MU-MISO NOMA downlink, and  propose two beam design methods together with corresponding user scheduling, by applying the transmitter-side non-linear processing technique, THP.
The contributions of this paper are summarized below:

$\bullet$  In Section \ref{sec:system_model}, we provide a framework for application of THP  to single-cell MU-MISO NOMA downlink systems. We show that by applying THP  to completely remove ICI to the strong users, the dimension of the beam design space  for the $k$-th cluster is increased to $N_t+1-k$ which is larger than that of simple ZF beam design $N_t+1-N_c$, where $N_t$ is the number of transmit antennas at the BS and $N_c$ is the number of clusters. The increased design freedom can be exploited to mitigate ICI to the weak users.

$\bullet$  In Section \ref{subsec:user_scheduling} , we propose a user scheduling algorithm together with
a greedy sequential beam design method by considering the rates of THP-aided MU-MISO NOMA. The proposed user scheduling algorithm first selects the strong users based on the semi-orthogonal user selection (SUS) algorithm \cite{Yoo&Goldsmith:06Jsel} and then selects the weak users sequentially with the beams designed in a sequential greedy manner.

$\bullet$  In Section \ref{subsec:joint_design}, we solve  the joint problem of beam redesign and power allocation  after user selection to further improve the performance over the sequential greedy beam design method.
This  joint optimization problem reduces to a non-convex problem. We propose an efficient algorithm to solve this joint optimization problem based on SCA and prove that the algorithm converges to a stationary point of the joint optimization problem.

Numerical results show that the proposed methods in this paper yield noticeable gain as compared to the existing methods for MU-MISO NOMA downlink.

\subsection{Notation and Organization}
\label{subsec:Notation}

We will use standard notations in this paper. Vectors and matrices are written in boldface with matrices in capitals. All vectors are column vectors. For a matrix $\Abf$, $\Abf^T$,
$\Abf^H$, $\Abf^{-1}$, and $\mathrm{Tr}(\Abf)$  indicate the transpose, conjugate transpose, inverse, and trace  of $\Abf$, respectively.
  $\Cc(\Abf)$ and $\Cc^\perp(\Abf)$ denote the
linear subspace spanned by the columns of $\Abf$ and its orthogonal
complement, respectively.
 $\Pibf_\Abf$ and
$\Pibf_\Abf^\bot$ are the projection matrices to $\Cc(\Abf)$ and
$\Cc^\perp(\Abf)$,
 respectively.
 $[\abf_1,\cdots,\abf_n]$ denotes the matrix composed of column vectors $\abf_1,\cdots,\abf_n$.
 $||\abf||$ represents the 2-norm of vector $\abf$.
 $\Ibf_n$ and $\mathbf{O}$ denote the $n\times n$ identity matrix (the subscript is omitted when unnecessary) and  all-zero matrix with proper size, respectively. For a random vector $\xbf$, $\Ebb\{\xbf\}$ denotes the expectation of $\xbf$, and $\xbf\sim\mathcal{CN}(\mubf,\Sigmabf)$ means that $\xbf$ is circularly-symmetric complex Gaussian-distributed with mean vector $\mubf$ and covariance matrix $\Sigmabf$. $\imath:=\sqrt{-1}$.

The remainder of this paper is organized as follows. In
Section II, the system model and preliminaries are described.
In Section III, the proposed method for user scheduling, beam design and power allocation for THP-aided MU-MISO NOMA systems is presented.  Numerical
results are provided in Section IV, followed by conclusions in
Section V.

\section{System Model}
\label{sec:system_model}

In this paper, we consider a single-cell MU-MISO NOMA downlink system consisting of a BS with $N_t$ transmit antennas and $K_{tot}$ single-antenna users.  We assume the following for our system model:

\vspace{0.3em}

\textit{A.1 (User Partition):} We assume that the total $K_{tot}$ users in the system  are partitioned into two user sets, $\Kc_1$ and $\Kc_2$, according to their channel strength, as in \cite{Choi15COM,Seo&Sung:18SP}. $\Kc_1$ is  the set of users with strong channels and  $\Kc_2$ is the set of users with weak channels. The cardinality of each set is given by $|\Kc_1|=|\Kc_2|=K_{tot}/2$.

\textit{A.2 (User Scheduling and Clustering):}
Taking signalling overhead and SIC error propagation into account,  we assume that  two users are grouped in each cluster as in \cite{Chen16:TSP,Kim13:MILCOM,Liu15:ICCW,Sayed17:WD}.  We assume that  $N_c~(\le N_t)$ clusters are constructed in total and  each cluster is composed of one strong-channel user (simply strong user) from $\Kc_1$ and one weak-channel user (simply weak user) from $\Kc_2$. In each cluster, the strong user performs SIC before decoding its own message, and the weak  user decodes its own data by treating the interference from the strong user as noise. For each cluster, we will refer to the strong  user as User 1 and the weak user as User 2. The details of user scheduling and grouping will be presented in Section \ref{sec:scheduling}.

\textit{A.3 (Spatial Multiplexing):}  To implement spatial multiplexing on top of two-user superposition coding in MU-MISO  NOMA, we assume that two users in each cluster are multiplexed in the power domain
with superposition coding and SIC as mentioned in Assumption \textit{A.2}, while multiple clusters are multiplexed in the spatial
domain by inter-cluster beamforming. To do so, we assume that a beam vector is assigned to each cluster and the strong and weak users in each cluster share the beam vector assigned to the cluster.
Under this
assumption, the transmit signal $\xbf$ of the BS for one scheduling
interval is given by
\begin{eqnarray} \label{eq:xbffirst}
    \xbf &=& \sum_{k=1}^{N_c} \wbf_k  \tilde{x}_k(\{d_{k1},d_{k2},p_{k1},p_{k2}\}_{k=1}^{N_c}),
\end{eqnarray}
where  $\wbf_k$ is the $N_t \times 1$ beam vector assigned to Cluster $k$ such that $||\wbf_k||^2 =1$, and $\tilde{x}_k(\{d_{k1},d_{k2},p_{k1},p_{k2}\}_{k=1}^{N_c})$ is the scalar signal of Cluster $k$ generated from the data symbols and the power values $\{d_{k1},d_{k2},p_{k1},p_{k2}, k=1,\cdots,N_c\}$. Here, $d_{k1}$ and $d_{k2}$ are the data symbols of Users 1 and 2, respectively, and $p_{k1}$ and $p_{k2}$ are the transmit power values assigned to Users 1 and 2, respectively, such that
\begin{equation}
p_{k1}+p_{k2}\le p_k,  ~~~\sum_{k=1}^{N_c}p_k\le P,
\end{equation}
where $p_k$ is the transmit power assigned to Cluster $k$ and $P$  is the total transmit power.

\textit{A.4 (Symbol Modulation):}
We assume that the user data symbols $d_{k1}$ and $d_{k2}$ are generated from  $M$-ary quadrature-based modulation such as $M$-ary amplitude modulation ($M$-QAM) or $M$-ary phase shift keying ($M$-PSK), and assume that $\Ebb\{|d_{k1}|^2\}=\Ebb\{|d_{k2}|^2\}=1$.

\vspace{0.5em}

Under the above assumptions, the received signals of Users 1 and 2 in Cluster $k$ are given by
\begin{align}
y_{k1} &= \hbf_{k1}^H\wbf_k\tilde{x}_k + \sum_{j\ne k}\hbf_{k1}^H\wbf_j\tilde{x}_j + n_{k1}, \label{eq:sysmodelrec1}\\
y_{k2} &= \hbf_{k2}^H\wbf_k\tilde{x}_k + \sum_{j\ne k}\hbf_{k2}^H\wbf_j\tilde{x}_j + n_{k2}  \label{eqA:sysmodelrec2}
\end{align}
where $y_{k1}$ and $y_{k2}$ are the received signals of Users 1 and 2 in Cluster $k$, $\hbf_{k1}$ and $\hbf_{k2}$ are the $N_t\times1$ channel vectors from the BS to Users 1 and 2 in Cluster $k$, and $n_{k1}$ and $n_{k2}$ are the additive white Gaussian noise (AWGN) at Users 1 and 2 in Cluster $k$ from distribution $\Cc\Nc(0,\sigma^2)$, respectively.
Here, the dependence of the cluster scalar signal $\tilde{x}_k$ on $\{d_{k1},d_{k2},p_{k1},p_{k2}\}$ is not shown for notational simplicity.
Note that the last two terms in each of
the right-hand sides (RHSs) of \eqref{eq:sysmodelrec1} and
\eqref{eqA:sysmodelrec2} are ICI and AWGN.

A widely-considered   way to combine MU-MISO with  NOMA  is to use  ZF inter-cluster beamforming for design of  $\wbf_1,\cdots,\wbf_{N_c}$  and to design the cluster scalar signal $\tilde{x}_k$  as \cite{Liu15:ICCW,Chen16:Access,Sayed17:WD,Seo&Sung:18SP}
\begin{equation}  \label{eq:ZFsymboldesign}
\tilde{x}_k(\{p_{k1},p_{k2},d_{k1},d_{k2}\}_{k=1}^{N_c}) = \sqrt{p_{k1}}d_{k1}+\sqrt{p_{k2}}d_{k2}.
\end{equation}
ZF beamforming is simple and effective to eliminate other user interference.   However, when it is applied to remove the ICI in MU-MISO NOMA,  it cannot remove the ICI for all users due to lack of spatial dimensions.  That is, in the case of $N_c=N_t$ with two-user grouping, we have $2N_t$ users but only $N_t$ transmit antennas. Hence, many researchers proposed using ZF beamforming to eliminate the ICI at the strong users \cite{Liu15:ICCW,Chen16:Access,Sayed17:WD,Seo&Sung:18SP}. This is because
 in NOMA, two users in each cluster are chosen so that the strong user has a high signal-to-noise ratio (SNR) channel, whereas the weak user is noise-limited. The high SNR channel is maintained by SIC for the strong user. On the other hand,  interference is allowed to the weak user with high noise anyway but high power is assigned to the weak user to boost its signal-to-interference-plus-noise ratio (SINR). Hence, to be consistent with this design principle of NOMA, ZF beamforming is applied to eliminate ICI at the strong users. In this case,
  the beam vector for Cluster $k$  is given by a unit-norm vector as
\begin{eqnarray}  \label{eq:ZFbeamVector}
    \wbf_k \in \Cc^\perp(\Hbf_1^{-k}),
\end{eqnarray}
where
\begin{equation} \label{eq:Hbf1mk}
\Hbf_1^{-k} := [\hbf_{11}, \cdots, \hbf_{k-1,1}, \hbf_{k+1,1},\cdots, \hbf_{N_c1}].
\end{equation}
 With the ZF beamforming vectors $\wbf_1,\cdots,\wbf_{N_c}$, we have $\hbf_{k1}^H \wbf_j =0, ~~\forall j \ne k$, and the ICI term in the received signal \eqref{eq:sysmodelrec1} of the strong user disappears.

However, the limitation of this ZF inter-cluster beamforming for MU-MISO NOMA is that it eliminates the design freedom for the beam vectors $\wbf_1,\cdots,\wbf_{N_c}$. In the case of $N_t=N_c$, the orthogonal space of the linear space spanned by the columns of
the matrix $\Hbf_1^{-k}$ in \eqref{eq:Hbf1mk} has  only one dimension almost surely for independently realized channel vectors $\hbf_{11},\cdots,\hbf_{N_c 1}$, and thus  the beam vector $\wbf_k$ for Cluster $k$ is predetermined by the channel vectors. Hence, we do not have control over $\wbf_k$ and the ICI at the weak user in \eqref{eqA:sysmodelrec2} is controlled only by user selection.  However, user selection alone has limitation in handling the ICI to the weak user.

To overcome this limitation of the ZF inter-cluster beamforming\footnote{One can consider minimum mean-square error (MMSE) inter-cluster beamforming to yield better performance at low SNR. However, MMSE beamforming converges to ZF beamforming at high SNR and the issue of the beam space restriction does not change with MMSE inter-cluster beamforming. That is, when the channel vectors $\hbf_{11},\cdots,\hbf_{N_c1}$ are given, the MMSE beam vectors are determined.}, in this paper we adopt THP at the BS to provide extra freedom to the design of  the inter-cluster beamforming vectors $\wbf_1,\cdots,\wbf_{N_c}$.  The application of THP is possible in a MU-MISO NOMA BS since the BS knows all data symbols for all downlink users. In the below, we briefly summarize the basic idea of THP and explain how THP can be applied to the  MU-MISO NOMA downlink.  Based on this, we will proceed to user scheduling and beam design in Section \ref{sec:scheduling}.

\subsection{Preliminary: Tomlinson-Harashima Precoding}
\label{subsec:THP}

THP is a nonlinear precoding technique which eliminates other user interference from the transmitter side based on channel state information at the transmitter (CSIT) like dirty paper coding (DPC), but it is a practical and usable technique\cite{Harashima&Miyakawa:72TCOM,Tomlinson:71EL,Windpassinger&Fischer&Vencel&Huber:04WC}. To explain THP in MU-MISO downlink,  let us consider a single-cell MU-MISO system with a BS with $N_t$ transmit antennas and $N_c(\le N_t)$ single-antenna users. The $N_c\times 1$ received signal vector $\ybf$ composed of the  received signals $y_1,\cdots,y_{N_c}$ of the $N_c$ users is given by
\begin{align}
\ybf = \Hbf^H\sbf + \nbf
\end{align}
where $\ybf=[y_1,y_2,\cdots,y_{N_c}]^T$ is the $N_c\times1$ received signal vector with $y_k$ being the received signal of the $k$-th user; $\Hbf = [\hbf_1,\hbf_2,\cdots,\hbf_{N_c}]$ is the $N_t\times N_c$ channel matrix  with $\hbf_k$ being the $N_t\times1$ MISO channel vector from the BS to the $k$-th user;  $\sbf$ is the $N_t\times 1$ transmit signal vector; and $\nbf=[n_1,n_2,\cdots, n_{N_c}]^T$ is the $N_c\times1$ noise vector with $n_k\stackrel{i.i.d.}{\sim}\Cc\Nc(0,\sigma^2)$. Note that the $N_t\times N_c$ matrix $\Hbf$ is a square or tall matrix since $ N_t \ge N_c$. By applying QR decomposition to $\Hbf$, we have
\begin{align}
\Hbf&=\Qbf\Rbf, ~~~~\Hbf^H = \Rbf^H\Qbf^H = \Lbf\Qbf^H,
\end{align}
where $\Lbf(=\Rbf^H)$ is an $N_c\times N_c$ lower-triangular matrix and $\Qbf$ is an $N_t\times N_c$ matrix whose column vectors $\{\qbf_k,k=1,\cdots,N_c\}$ are orthogonal to each other with $\|\qbf_k\|^2=1$ for $k=1,\cdots,N_c$. Let $\sbf=\Qbf\tilde{\sbf}$, where $\tilde{\sbf}=[\tilde{s}_1,\cdots,\tilde{s}_{N_c}]^T$ is an $N_c\times 1$ effective transmit signal vector and  $\tilde{s}_k$ is the  transmit signal for the $k$-th user. (Note  that in this case  $\Qbf$ is the transmit beamforming matrix and $\qbf_k$ is the beam vector for the $k$-th user carrying $\tilde{s}_k$.)  Then, the received signal vector $\ybf$ can be rewritten as
\begin{align}
\ybf &= \Hbf^H\sbf + \nbf  = \Lbf\Qbf^H\Qbf\tilde{\sbf} + \nbf = \Lbf\tilde{\sbf} + \nbf \label{eq:THP_beam}
\end{align}
and the corresponding received signal at the $k$-th user is given by
\begin{align}  \label{eq:THP1}
y_k &= l_{kk}\tilde{s}_k + \sum_{j=1}^{k-1}l_{kj}\tilde{s}_{j} + n_k,~~k=1,\cdots,N_c,
\end{align}
where $l_{kj}$ is the element of $\Lbf$ at the $k$-th row and $j$-th column, the first term in the RHS of \eqref{eq:THP1} is the desired signal of the $k$-th user, and the second term in the RHS of \eqref{eq:THP1} is the   interference  from the first to $(k-1)$-th users to the $k$-th user.

\begin{figure}[t]
\centerline{
\psfragscanon
\psfrag{A/2}[t][t]{\small $A/2$}
\includegraphics[scale=0.7]{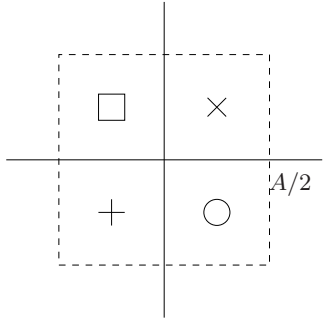}}
\caption{4-QAM constellation points}
\label{fig:4QAMconst}
\end{figure}

\begin{figure}[t]
\centerline{
\includegraphics[scale=0.45]{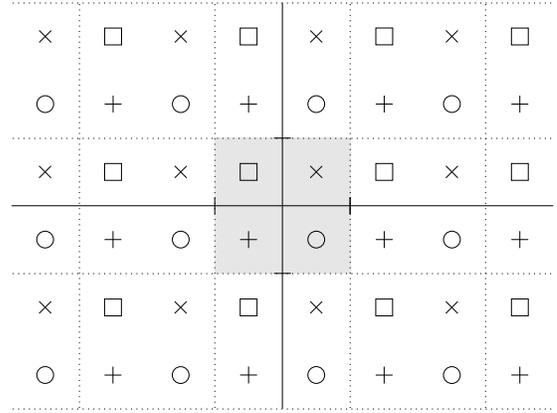}
}
\caption{Expanded constellation points of the 4-QAM in Fig. \ref{fig:4QAMconst} }
\label{fig:const_expanded}
\end{figure}

THP exploits the fact that in digital communication the  symbols for each user are modulated symbols existing only on a certain modulation constellation.  Suppose that the actual data symbol  $d_k$ carried in the transmit signal $\tilde{s}_k$ is  from the set of $M$-ary quadrature-based modulation constellation points. For the purpose of explanation, consider the constellation $\{(+\frac{A}{4},+\frac{A}{4}),(+\frac{A}{4},-\frac{A}{4}),(-\frac{A}{4},+\frac{A}{4}),(-\frac{A}{4},-\frac{A}{4})\}$ of 4-QAM here, as shown in Fig. \ref{fig:4QAMconst}.
In order to eliminate interference $\sum_{j=1}^{k-1}l_{kj}\tilde{s}_{j}$ in the received signal of the $k$-th user \eqref{eq:THP1}, THP subtracts the interference and uses modulo operation  in a sequential manner at the transmitter side as follows:\cite{Windpassinger&Fischer&Vencel&Huber:04WC}
\begin{align}
\tilde{s}_1 &= d_1  \label{eq:THPencoding1}\\
\tilde{s}_2 &= \mathrm{mods}_A\left(d_2 - \frac{l_{21}}{l_{22}}\tilde{s}_1\right)\\
&\vdots  \nonumber\\
\tilde{s}_k &= \mathrm{mods}_A\left(d_k - \sum_{j<k}\frac{l_{kj}}{l_{kk}}\tilde{s}_j\right),  \label{eq:ModuloLast}
\end{align}
where $\mathrm{mods}_A(x)$ is the symmetric modulo operation that returns the remainder  of $x$ after division by $A$ such that  $\mathrm{real}(\mathrm{mods}_A(x))\in[-A/2,A/2)$ and $\mathrm{imag}(\mathrm{mods}_A(x))\in[-A/2,A/2)$.  This modulo operation  maintains the transmit power of the signal. Then, from \eqref{eq:THP1} the received signal at the $k$-th user is given by
\begin{align}
y_k &= l_{kk}\tilde{s}_k + \sum_{j<k}l_{kj}\tilde{s}_j + n_k\\
    &= l_{kk}(d_k + c_R A + \imath c_I A) + n_k,  \label{eq:ModuloLast2}
\end{align}
where $c_R$ and $c_I$ are some integers decided by the modulo operation.  Here,  \eqref{eq:ModuloLast2} is valid since $\tilde{s}_k =\mathrm{mods}_A\left(d_k - \sum_{j<k}\frac{l_{kj}}{l_{kk}}\tilde{s}_j\right)$ in \eqref{eq:ModuloLast} can be expressed as $\tilde{s}_k=d_k - \sum_{j<k}\frac{l_{kj}}{l_{kk}}\tilde{s}_j + c_R A + \imath c_I A$ for some integers $c_R$ and $c_I$.
The processing at the receiver side is rather simple. By dividing the received signal $y_k$ by the effective gain $l_{kk}$ of the $k$-th user, we can decode the data symbol $d_k$ by using infinitely expanded constellation points shown in Fig. \ref{fig:const_expanded}. That is, the normalized received signal $y_k/l_{kk}$ is located in a certain shifted box in Fig. \ref{fig:const_expanded} and the demodulation of the data symbol $d_k$ is performed in that shifted box.
One can see that each user has no other user interference and the power of the transmit signal $\tilde{s}_k$ is  similar to the power of the original $M$-QAM constellation  since $\tilde{s}_k$ is contained in the boundary of the original  $M$-QAM constellation shown in Fig. \ref{fig:4QAMconst} by the modulo-operation.

In \eqref{eq:THP_beam}, to apply THP to MU-MISO downlink,   we  used $\Qbf=[\qbf_1,\cdots,\qbf_{N_c}]$ obtained from QR decomposition of $\Hbf$ as the transmit beamforming matrix   to make the resulting effective channel matrix as a low-triangular matrix. In fact, to make the resulting effective channel matrix as a low-triangular matrix, we can use any beamforming matrix $\Wbf=[\wbf_1~\cdots~\wbf_{N_c}]$ satisfying the following condtion:
\begin{align}
\wbf_k \in \Cc^\perp([\hbf_1,\hbf_2,\cdots,\hbf_{(k-1)}]).
\end{align}
Hence, the design space for the first user's beam vector  $\wbf_1$ is the entire space ${\mathbb{C}}^{N_t}$, the design space for $\wbf_2$ is ${\mathbb{C}}^{N_t-1}$, and the design space for the $k$-th user's beam vector $\wbf_k$ is ${\mathbb{C}}^{N_t-k}$.
On the other hand, for the ZF inter-cluster beamforming, the design space for $\wbf_k$ is ${\mathbb{C}}^{N_t-N_c+1}$ regardless of $k$, as seen in  \eqref{eq:ZFbeamVector}. When $N_t=N_c$, the beam space for all $\wbf_k$ is one-dimensional and determined by the channel vectors in the inter-cluster ZF beamforming case. Thus, the dimension of the beam design space is much increased by adopting THP and we can exploit this beam design freedom in addition to user scheduling to control the ICI to the weak users and to enhance the overall performance in MU-MISO NOMA.

\subsection{Non-Orthogonal Multiple Access with THP}
\label{subsec:THP_NOMA}

In this subsection, we explain how to apply THP to the considered MU-MISO NOMA downlink system.
As aforementioned, in the considered NOMA system,   two users in each cluster are chosen so that the strong user has high-quality channel, whereas the weak user is noise-limited. The high-quality channel for the strong user  should be  maintained for proper operation of NOMA. To be consistent with this design principle of NOMA, we apply THP and design inter-cluster beam vectors so that ICI is eliminated for  the strong users.  From \eqref{eq:xbffirst} and \eqref{eq:sysmodelrec1}, the matrix model for the received signals at the strong users at all clusters is given by
\begin{align}
\ybf_1 &= \Hbf_1^H \xbf + \nbf_1 =\Hbf_1^H [\wbf_1,\cdots,\wbf_{N_c}]\left[
\begin{array}{c}
\tilde{x}_1\\
\vdots\\
\tilde{x}_{N_c}
\end{array}
\right]+\nbf_1,
\end{align}
where $\ybf_1=[y_{11},y_{21},\cdots,y_{N_c1}]^T$, $\Hbf_1=[\hbf_{11},\cdots,\hbf_{N_c1}]$,  and $\nbf_1=[n_{11},n_{21},\cdots,n_{N_c 1}]^T$.
Based on the discussion in Section \ref{subsec:THP}, to apply sequential THP
\eqref{eq:THPencoding1} - \eqref{eq:ModuloLast}
to the strong users, we require the inter-cluster beam vectors $\wbf_k$, $k=1,\cdots,N_c$ to satisfy the following constraint:
\begin{equation} \label{eq:THP_NOMA_beam}
\wbf_k \in \Cc^\perp(\Hbf_1^{<k}),
\end{equation}
where
\begin{equation}
\Hbf_1^{<k} := [\hbf_{11}~\hbf_{21}~\cdots~\hbf_{(k-1)1}]
\end{equation}
 and $\hbf_{k1}$ is the channel vector from the BS to the strong user in Cluster $k$. Thus, the dimension of the beam design space for the $k$-th cluster is $N_t-k+1$, whereas that of the ZF inter-cluster beamforming is
 $N_t-N_c+1$.   With the inter-cluster beam vectors $\wbf_1,\cdots,\wbf_{N_c}$ satisfying \eqref{eq:THP_NOMA_beam}, we have the signal model for the strong user of Cluster $k$ as
 \begin{equation}
 y_{k1} = \hbf_{k1}^H\wbf_k\tilde{x}_k + \sum_{j=1}^{k-1}\hbf_{k1}^H\wbf_j\tilde{x}_{j} + n_{k1},
 \end{equation}
which is in the same form as   \eqref{eq:THP1}. Thus, THP   \eqref{eq:THPencoding1} - \eqref{eq:ModuloLast} is applied sequentially to the transmit signal $\tilde{x}_k$ as
\begin{align}
\tilde{x}_k &= \mathrm{mods}_B\left(x_k - \sum_{j<k}\frac{\hbf_{k1}^H\wbf_j}{\hbf_{k1}^H\wbf_k}\tilde{x}_j\right),\label{eq:THP_NOMA_signal}
\end{align}
where $B$ is  the modulo operation factor. Note that THP encoding requires the CSIT of the strong users only. Since we implement NOMA, the signal $x_k$ intended for Cluster $k$ is designed by superposition coding as
\begin{align}
x_k &= \sqrt{p_{k1}}d_{k1} + \sqrt{p_{k2}}d_{k2},\\
p_k &= p_{k1} + p_{k2},~~ \sum_{k=1}^{N_c}p_k \le P,
\end{align}
where $d_{k1}$ and $d_{k2}$ are the modulation data symbols for the strong user and the weak user of the $k$th cluster, respectively, $p_{k1}$ and $p_{k2}$ are transmit signal power for the strong user and the weak user of Cluster $k$, respectively, $p_k$ is the power of Cluster $k$, and $P$ is the total transmit power of the BS. Note that the modulo-operation in \eqref{eq:THP_NOMA_signal} should be performed with an appropriate  $B$  by considering the boundary of the super-imposed constellation points. Based on \eqref{eq:THP_NOMA_beam} and \eqref{eq:THP_NOMA_signal}, the received signals
 \eqref{eq:sysmodelrec1} and \eqref{eqA:sysmodelrec2}
of the strong and weak users of Cluster $k$ are given by
\begin{align}
y_{k1}       &= \hbf_{k1}^H\wbf_kx_k + n_{k1} \label{eq:THP_NOMA_strong111}\\
y_{k2} &= \hbf_{k2}^H\wbf_k\tilde{x}_k + \sum_{j\ne k}\hbf_{k2}^H\wbf_j\tilde{x}_j + n_{k2} \nonumber \\
       &= \hbf_{k2}^H\wbf_kx_k + \sum_{j<k}\left(\hbf_{k2}^H\wbf_j - \frac{\hbf_{k2}^H\wbf_k}{\hbf_{k1}^H\wbf_k}\hbf_{k1}^H\wbf_j\right)\tilde{x}_j \nonumber\\
       &~~~~~~~~~~~~~~~ + \sum_{j>k}\hbf_{k2}^H\wbf_j\tilde{x}_j + n_{k2}, \label{eq:THP_NOMA_weak111}
\end{align}
where the shifting constants $c_R B + \imath c_I B$  associated with the modulo-operation are omitted in \eqref{eq:THP_NOMA_strong111} and \eqref{eq:THP_NOMA_weak111} for  simplicity under the assumption that  the strong user in Cluster $k$ decodes $x_k=\sqrt{p_{k1}}d_{k1} + \sqrt{p_{k2}}d_{k2}$ based on $y_{k1}/(\hbf_{k1}^H\wbf_k)$ and the weak user in Cluster $k$ decodes $x_k=\sqrt{p_{k1}}d_{k1} + \sqrt{p_{k2}}d_{k2}$ based on $y_{k2}/(\hbf_{k2}^H\wbf_k)$  by using infinitely expanded constellation points.  Precisely, the strong user decodes $d_{k2}$ first and then decodes $d_{k1}$ from the interference-cancelled signal $x_k - \sqrt{p_{k2}}d_{k2}$, whereas the weak user decodes $d_{k2}$ in $x_{k}$ treating $d_{k1}$ as noise.
As seen in \eqref{eq:THP_NOMA_strong111} and \eqref{eq:THP_NOMA_weak111}, the ICI disappears at the strong user as in the case of ZF inter-cluster beamforming, but the ICI remains at the weak user. However, the components of the received signal at the weak user are different from those of the ZF inter-cluster beamforming case,
as seen in \eqref{eq:THP_NOMA_weak111}.
The corresponding rates of the strong user and the weak user in Cluster $k$ with the proposed inter-cluster beamforming, THP and SIC  are respectively given by\footnote{Here, we neglect the rate loss induced by modulation quantization from the Gaussian input signal. As the order of modulation increases, this quantization loss becomes small. Depending on the SNR, we can select the modulation order adaptively to approach the Gaussian-input rate.}
\begin{align}
R_{k1} &= \log_2\left(1+\frac{p_{k1}|\hbf_{k1}^H\wbf_k|^2}{\sigma^2}\right),\label{eq:THP_NOMA_rate11}\\
R_{k2} &= \log_2\left(1+\min\left\{\frac{p_{k2}|\hbf_{k1}^H\wbf_k|^2}{p_{k1}|\hbf_{k1}^H\wbf_k|^2+\sigma^2},\frac{p_{k2}|\hbf_{k2}^H\wbf_k|^2}{I_k + \sigma^2}\right\}\right) \label{eq:THP_NOMA_rate22}
\end{align}
where
\begin{align}
I_k &= p_{k1}|\hbf_{k2}^H\wbf_k|^2 + \sum_{j<k}p_j\left|\hbf_{k2}^H\wbf_j - \frac{\hbf_{k2}^H\wbf_k}{\hbf_{k1}^H\wbf_k}
{\hbf_{k1}^H\wbf_j}\right|^2\nonumber \\
&~~~~~~~~~~~~~~~~~~~ + \sum_{j>k}p_j|\hbf_{k2}^H\wbf_j|^2.\label{eq:THP_NOMA_interf33}
\end{align}
In computation of  the strong user rate $R_{k1}$ in \eqref{eq:THP_NOMA_rate11}, the interference from the weak user is not shown  since the strong user applies SIC to the interference from the weak user before decoding its own message. The weak user rate $R_{k2}$ is determined by two factors. First, the weak user's data should be decodable at the strong user before decoding the strong user's data at the strong user and the first term in the minimum in  \eqref{eq:THP_NOMA_rate22} represents this rate. Second, the weak user's data should be decodable at the weak user itself, while treating all other signals as noise, and the second term in the minimum in  \eqref{eq:THP_NOMA_rate22} represents this rate. Note that the first term in the RHS of \eqref{eq:THP_NOMA_interf33} is the interference from the strong user of the same cluster and the second and third terms in the RHS of \eqref{eq:THP_NOMA_interf33} are ICI. Note that by properly designing the beam vectors we can control the weak user interference $I_k$ in \eqref{eq:THP_NOMA_interf33} and consequently the weak user rate in \eqref{eq:THP_NOMA_rate22}.

\section{Proposed User Scheduling and Beam Design}
\label{sec:scheduling}

In the previous section, we explained  how to apply THP to MU-MISO NOMA, and derived the achievable rates of each cluster in the MU-MISO NOMA with THP. In this section, we now tackle the main problem of user scheduling, beam design and power allocation for THP-aided MU-MISO NOMA.
In the previous works in which   ZF inter-cluster beamforming  \eqref{eq:ZFbeamVector} is considered, the problem of beam design is simple  since the ZF constraint \eqref{eq:ZFbeamVector} determines the beam vectors, and only the problem of user scheduling and power allocation remains. On the contrary, in the case of the considered THP-aided MU-MISO NOMA, we have the further freedom of designing the inter-cluster beam vectors under the relaxed constraint \eqref{eq:THP_NOMA_beam} in addition to the freedom of user scheduling and power allocation.
There exist several optimality criteria based on the rates of the strong and weak users in \eqref{eq:THP_NOMA_rate11} and \eqref{eq:THP_NOMA_rate22}. One may consider the  maximization of the sum of strong and weak users' rates. However,  in the asymmetric channel case where NOMA is meaningful\footnote{In the case of two-user symmetric channels between strong and weak users, orthogonal multiple access (OMA) is optimal and it achieves the boundary of the capacity region\cite{Tse&Viswanath:book}.},
all cluster power would be allocated to the strong user  and this would make the weak user's rate zero,   if the criterion of maximization of the sum of strong and weak users' rates were adopted.
Hence, one reasonable optimality criterion in this case is to maximize the sum of  weak users' rates while guaranteeing certain target rates for the strong users.  Thus, under the considered THP-aided MU-MISO NOMA, we consider the optimal  beam design and power allocation problem,  formulated as follows:

\vspace{0.5em}

\begin{problem}  \label{problem:joint_beam}
Given   total power $P$ and strong user target SNR parameter $\eta$,
maximize the sum of  weak users' rates, i.e.,
\begin{equation}  \label{eq:prob1_obj}
\mathop{\max}\limits_{R_{k2},\wbf_k,p_{k1},p_{k2},\forall k} ~ \sum_{k=1}^{N_c} R_{k2}
\end{equation}
subject to
\begin{align}
& \frac{|\hbf_{k1}^H\wbf_k|^2p_{k1}}{\sigma^2}\ge \eta\frac{P}{N_c}\frac{|\Pibf_{\Hbf_1^{<k}}^\perp \hbf_{k1}|^2}{\sigma^2},~\forall k \label{eq:prob1_R1}\\
& R_{k2} \le \log_2\left(1+\frac{|\hbf_{k1}^H\wbf_k|^2p_{k2}}{|\hbf_{k1}^H\wbf_k|^2p_{k1}+\sigma^2}\right),~\forall k \label{eq:prob1_R21}\\
& R_{k2} \le \log_2\left(1+\frac{|\hbf_{k2}^H\wbf_k|^2p_{k2}}{I_k + \sigma^2}\right),~\forall k \label{eq:prob1_R22}\\
& (\Hbf_1^{<k})^H\wbf_k = \mathbf{0},~\forall k  \label{eq:prob1_null}\\
& \|\wbf_k\|^2 \le 1,~\forall k \label{eq:prob1_norm}\\
& \sum_k (p_{k1} + p_{k2}) \le P, \label{eq:prob1_power}
\end{align}
where  $I_k$ is given by \eqref{eq:THP_NOMA_interf33}, $\{\hbf_{11},\cdots,\hbf_{N_c1}\}$ are the channel vectors of
the scheduled strong users, and  $\{\hbf_{12},\cdots,\hbf_{N_c2}\}$ are
the channel vectors of the scheduled weak users. (Note that in Problem \ref{problem:joint_beam}, $R_{k2}$ is used as a slack variable and the cost function is linear and hence convex in the overall optimization variables.)
Here, the condition \eqref{eq:prob1_R1} is to guarantee a certain target rate for each strong user, where the strong user rate is given by \eqref{eq:THP_NOMA_rate11}. Note that the condition is expressed in terms of SNR. The term $\frac{P}{N_c}{|\Pibf_{\Hbf_1^{<k}}^\perp \hbf_{k1}|^2}/{\sigma^2}$ in the RHS of \eqref{eq:prob1_R1} represents the nominal maximum SNR for the strong user when $\wbf_k$ is the matched-filtering beam to $\hbf_{k1}$ under the THP beam constraint   \eqref{eq:THP_NOMA_beam}.  Thus, the target SNR for each strong user is the $\eta$-fraction of this nominal maximum SNR. The conditions \eqref{eq:prob1_R21} and \eqref{eq:prob1_R22} implement \eqref{eq:THP_NOMA_rate22}.  The condition \eqref{eq:prob1_null} simply realizes the THP beam constraint   \eqref{eq:THP_NOMA_beam}.  The constraints \eqref{eq:prob1_norm} and \eqref{eq:prob1_power} are power constraints.
\end{problem}

\vspace{0.5em}

The   joint design problem of beam design, power allocation and user scheduling  under the optimality in Problem \ref{problem:joint_beam} is a complicated problem since the rates are dependent not only  on the beam design and power allocation but also on user scheduling.   To circumvent this difficulty, we apply a three-step approach to the complicated joint problem of  beam design, power allocation and user scheduling. The three steps are as follows:

{\textit{S.1)}} We first select $N_c$ strong users from the strong user set $\Kc_1$ by using the SUS algorithm \cite{Yoo&Goldsmith:06Jsel}.

{\textit{S.2)}} For the  set of  strong users obtained from Step {\textit{S.1)}}, we then select weak users one by one, while designing the cluster beam vector sequentially under the assumption that equal power is allocated to every cluster.

{\textit{S.3)}} Finally, with the selected strong and weak users we solve the problem of beam redesign  and power allocation to maximize the performance.

Although the proposed multi-step approach is not an optimal solution to the joint problem of  beam design, power allocation and user scheduling, it provides a tractable and efficient solution to the joint problem.  It will be shown in Section \ref{sec:numerical_results} that the proposed method yields noticeable gain over the existing ZF inter-cluster beamforming-based methods.  We explain the steps in detail below.

\subsection{User Scheduling via Sequential Greedy Beam Design}
\label{subsec:user_scheduling}

The steps {\textit{S.1}} and {\textit{S.2}} are basically user scheduling and grouping.  User scheduling begins with the selection of $N_c$ strong users from the strong user set $\Kc_1$ in Step {\textit{S.1}}. The initial selection of strong users based only on the channels simplifies the overall problem significantly and such initial separate selection of strong users based on the SUS algorithm was considered in other works as well \cite{Liu15:ICCW,Seo&Sung:18SP}. For the selection of strong users, even though we use beamforming and THP to eliminate the ICI to the strong users, the strong users with orthogonal channel vectors $\hbf_{11},\cdots,\hbf_{N_c1}$ are preferred. This is because in case of orthogonal channel vectors $\hbf_{11},\cdots,\hbf_{N_c1}$, by setting $\wbf_k=\frac{\hbf_{k1}}{||\hbf_{k1}||}$, there is no ICI and the gain of the resulting individual strong user communication channel is maximized. Furthermore, for NOMA it is better that  strong users have high quality channels, which translates to channels with large channel norms under rough orthogonality.  The  SUS algorithm  is suitable for this purpose because it
chooses users with orthogonal channels with large channel norms.
  We set the $N_c$ strong users returned by the SUS algorithm as the $N_c$ strong users in the $N_c$ clusters (one for each cluster).

\begin{algorithm}[t]
   \caption{\textbf{{User Scheduling with Sequential Beam Design}}}\label{THP:algorithm1}
    \begin{algorithmic}[1]
       \Step \textbf{S.1)} Strong user selection
       \State Run the SUS algorithm \cite{Yoo&Goldsmith:06Jsel} to select $N_c (\le N_t)$ users from the strong user set $\Kc_1$.
       \State Obtain the selected strong users' channels $\hbf_{11}, \cdots, \hbf_{N_c1}$.
       \Step \textbf{S.2)} Weak user selection
       \State \textbf{Initialization:}  Given information:
       $\gbf_1, \cdots,\gbf_{|\Kc_2|}$ (user channel vectors in $\Kc_2$), $P$ (total power), $\eta$ (strong user target SNR parameter).
       \State  $p_k = P/N_c$ and $\hat{p}_{k1}=\eta p_k, \forall k$
       \State $\Kc_2 \gets \{ 1, \ldots , K_{tot}/2 \}$  \Comment{the original weak user set}
       \State $\Sc_1 $ \Comment{the set of selected strong users from step S.1)}
       \State $\Sc_2 \gets \phi$ \Comment{the set of selected weak users}
       \State $[\hat{\wbf}_1,\cdots,\hat{\wbf}_{N_c}] = \left[\frac{\Pibf_{\Hbf_1^{<k}}^\perp \hbf_{11}}{||\Pibf_{{\Hbf}_1^{<k}}^\perp \hbf_{11}||}, \cdots, \frac{\Pibf_{\Hbf_{N_c}}^\perp \hbf_{N_c1}}{||\Pibf_{{\Hbf}_{N_c}^{<k}}^\perp \hbf_{N_c1}||} \right]$
       \State $\Wbf=[~]$,
       \State \textbf{Execution}:
       \For{$k=1$ to $N_c$}
                  \State \textbf{(S.2-1)} Compute the rate of every candidate weak user.
            \For{$u = 1$ to $|\Kc_2|$}
                  \begin{align}
                    \widehat{I}_u &= |\gbf_u^H \hat{\wbf}_k|^2\hat{p}_{k1} + \sum_{j<k}\left|\gbf_{u}^H{\wbf_j} - \frac{\gbf_{u}^H{\hat{\wbf}_k}}{\hbf_{k1}^H{\hat{\wbf}_k}}{\hbf_{k1}^H{\wbf_j}}\right|^2p_j\nonumber\\
                    &~~~~~~~~~~~~~~~~~~+ \sum_{j>k}|\gbf_{u}^H{\hat{\wbf}_j}|^2p_j,  \label{eq:estICIinAlgo}
                \end{align}
                \hspace{3em}where $\wbf_l$ is the $l$-th column of $\Wbf$.
                \State Compute $\hat{R}_u$ based on \eqref{eq:THP_NOMA_rate22} with  $\widehat{I}_u$ and $\hat{\wbf}_k$.
            \EndFor
                \vspace{0.5em}
            \State \textbf{(S.2-2) } Select the weak user for cluster $k$.
            \State  $u^* = \mathop{\arg\max}\limits_{u \in \Kc_2} \hat{R}_{u}$
            \State  $\Sc_2 \gets \Sc_2 \cup \{u^*\}$ and $\hbf_{k2}\leftarrow\gbf_{u^*}$
            \State  $\Kc_2 \gets \Kc_2 \backslash \{u^*\}$
               \vspace{0.5em}
            \State \textbf{(S.2-3)} Greedy beam design:
            \State Design $\wbf_k$ by solving Problem \ref{problem:sequetial_beam}.
            \State $\Wbf\leftarrow [\Wbf, \wbf_k]$
       \EndFor
\end{algorithmic}
\end{algorithm}

After strong user selection,  weak users are sequentially selected for each cluster from the weak user set $\Kc_2$ by  sequentially designing the beam vector for each cluster in a greedy manner under the assumption of equal cluster power allocation. The proposed algorithm is summarized in Algorithm \ref{THP:algorithm1}.  The flow of Algorithm  \ref{THP:algorithm1} is similar to that of the user scheduling algorithm in \cite{Seo&Sung:18SP}. However, Algorithm  \ref{THP:algorithm1} has several distinctive features relevant to the considered THP-aided MU-MISO NOMA. Note that Algorithm \ref{THP:algorithm1} chooses the weak user sequentially from cluster 1 to $N_c$ as seen in Lines 11 to 23 in Algorithm \ref{THP:algorithm1}.  At the time when the weak user is selected and the beam vector is designed for Cluster $k$, the weak users and beam vectors for Cluster $k+1,\cdots,N_c$ are not determined yet. However, the beam vector information for Cluster $k+1,\cdots,N_c$ is required to compute the ICI in \eqref{eq:THP_NOMA_interf33}, which is required in turn to compute the weak user rate in \eqref{eq:THP_NOMA_rate22}. Hence, we use the matched-filtering beams under the THP beam constraint \eqref{eq:THP_NOMA_beam} as the estimates of the undetermined beams for Clusters $k+1,\cdots,N_c$, as shown in Line 8 of Algorithm \ref{THP:algorithm1}. With the already designed beams for Clusters $j<k$ and the beam estimates for Clusters $j \ge k$, the ICI can be estimated as  \eqref{eq:estICIinAlgo}, the candidate weak user rate is estimated and   the weak user is selected for Cluster $k$. Then, the beam vector for Cluster $k$ is designed by solving Problem \ref{problem:sequetial_beam}, which is a greedy version of our design criterion in Problem 1:

\vspace{0.5em}
\begin{problem}  \label{problem:sequetial_beam}
\begin{align}
\begin{array}{cl}
\mathop{\max}\limits_{R_{k2},\wbf_k,p_{k1},p_{k2}} &  R_{k2}\\
\mbox{subject to} & \frac{|\hbf_{k1}^H\wbf_k|^2p_{k1}}{\sigma^2}\ge \eta\frac{P}{N_c}\frac{|\Pibf_{\Hbf_{1}^{<k}}^\perp \hbf_{k1}|^2}{\sigma^2}\\
& R_{k2} \le \log_2\left(1+\frac{|\hbf_{k1}^H\wbf_k|^2p_{k2}}{|\hbf_{k1}^H\wbf_k|^2p_{k1}+\sigma^2}\right)\\
& R_{k2} \le \log_2\left(1+\frac{|\hbf_{k2}^H\wbf_k|^2p_{k2}}{I^\prime_k + \sigma^2}\right)\\
& (\Hbf_1^{<k})^H\wbf_k = \mathbf{0}  \\
& \|\wbf_k\|^2 \le 1\\
& p_{k1} + p_{k2} \le P/N_c
\end{array}
\end{align}
where
\begin{align}
I^\prime_k &= |\hbf_{k2}^H\wbf_k|^2p_{k1} + \sum_{j<k}\left|
\hbf_{k2}^H\wbf_j-
\frac{\hbf_{k2}^H\wbf_k}{\hbf_{k1}^H\wbf_k}  \hbf_{k1}^H\wbf_j \right|^2 \frac{P}{N_c}\nonumber \\
&~~~~~~~~~~~~~~~~~~~ + \sum_{j>k}|\hbf_{k2}^H{\hat{\wbf}_j}|^2\frac{P}{N_c}.\label{eq:estimated_interf}
\end{align}
\end{problem}
Note that Problem  \ref{problem:sequetial_beam} is a non-convex problem. How to solve Problem  \ref{problem:sequetial_beam} is discussed in the next subsection.

\subsection{Joint Beam Design and Cluster Power Allocation}
\label{subsec:joint_design}

In the previous subsection, we considered  user scheduling with greedy sequential beam design  under the assumption of equal cluster power allocation.  Although the sequentially designed beams can be used directly together with the strong and weak users selected by Algorithm \ref{THP:algorithm1}, it is not optimal under the criterion considered in Problem 1.
In this section,  we consider the joint  beam redesign and power allocation problem, formulated in Problem 1, with the scheduled strong and weak users obtained by Algorithm \ref{THP:algorithm1}.

Problem  \ref{problem:joint_beam} is a non-trivial non-convex  optimization problem due to the three non-convex constraints \eqref{eq:prob1_R1}, \eqref{eq:prob1_R21} and \eqref{eq:prob1_R22}.  Especially, the ICI term $I_k$ in the decoding of $x_2$ at the weak user has a complicated form as seen in \eqref{eq:THP_NOMA_interf33}, and hence  it is not straightforward how to approach Problem \ref{problem:joint_beam}. Simple iterative methods may not even guarantee convergence.
To solve Problem \ref{problem:joint_beam}, we resort to SCA \cite{Scutari&Facchinei&Lampariello:17SP}. SCA is a  method to solve a non-convex optimization problem by iteratively solving  properly-constructed approximating convex optimization problems for the original non-convex optimization.
It is known that if the approximating convex optimization problems satisfy certain conditions, a stationary point of the original non-convex optimization problem can be obtained by SCA \cite{Scutari&Facchinei&Lampariello:17SP}. Below, we summarize the recent result about SCA in \cite{Scutari&Facchinei&Lampariello:17SP} as a theorem relevant to   Problem  \ref{problem:joint_beam}.

\vspace{0.5em}

{\it Theorem 1 \cite{Scutari&Facchinei&Lampariello:17SP}:}
Consider a non-convex optimization problem $\Pc$ with a convex objective function $c_0(\abf)$ and non-convex constraints $c_l(\abf)\le 0,~ 1 \le l \le L$,  where $\abf$ is the optimization variable of $\Pc$. Let $\tilde{c}_l(\abf,\abf^{(i)}) \le 0$ be
the convex constraints approximating the original non-convex constraints $c_l(\abf) \le 0$,  where $\abf^{(i)}$ is the point for approximation at iteration $i$, and let the approximating convex optimization problem $\Pc^{(i)}$ at iteration $i$ be given by the same cost function $c_0(\abf)$ with $\tilde{c}_l(\abf,\abf^{(i)})\le 0$ replacing $c_l(\abf)\le 0$. Then, if the conditions C1 - C8 below  are satisfied, the NOVA algorithm \cite{Scutari&Facchinei&Lampariello:17SP} with step-size $\gamma^{(i)}$ satisfying
\begin{align}  \label{eq:stepsizeTheo1}
0<\inf_i\gamma^{(i)}\le\sup_i\gamma^{(i)}\le \max_i\gamma^{(i)}\le 1
\end{align}
yields a stationary point of the problem $\Pc$ as a limit point.
\begin{enumerate}[C1)]
\item $c_l$ are continuously differentiable on $\Uc$, where $\Uc$ is a closed and convex set containing the feasible set $\Ac$ for $\Pc$;
\item $\tilde{c}_l(\bullet;\abf)$ is convex on $\Uc$ for all $\abf\in\Ac$.
\item $\tilde{c}_l(\abf;\abf)=c_l(\abf)$, for all $\abf\in\Ac$;
\item $c_l(\abf)\le \tilde{c}_l(\abf,\bbf)$ for all $\abf\in\Uc$ and $\bbf\in\Ac$;
\item $\tilde{c}_l(\bullet;\bullet)$ is continuous on $\Uc\times\Ac$;
\item $\nabla_\bbf c_l(\abf) = \nabla_\bbf \tilde{c}_l(\abf;\abf)$ for all $\abf\in\Ac$;
\item $\nabla_\bbf \tilde{c}_l(\bullet;\bullet)$ is continuous on $\Uc\times \Ac$.
\item The approximating convex optimization problem $\Pc^{(i)}$ satisfies Slater's condition.
\end{enumerate}
 Here, $\nabla_\bbf\tilde{c}_l(\abf;\abf)$ denotes the partial gradient of $\tilde{c}_l$ w.r.t. the its first argument evaluated at $\abf$.
\begin{algorithm}[t]
   \caption{\textbf{{The NOVA Algorithm}}\cite{Scutari&Facchinei&Lampariello:17SP}}\label{NOVA_algorithm}
    \begin{algorithmic}
        \State \hspace{-1.4em} \textbf{Data}:   Step size $\gamma^{(i)}\in(0,1]$, initial point $\abf^{(0)}\in\Ac$. Set $i=0$.
        \Step \textbf{1)} If $\abf^{(i)}$ is a stationary point of $\Pc$ (i.e., the cost $c_0$ does not change any further), stop.
        \Step \textbf{2)} Set $\hat{\abf}(\abf^{(i)})$ as the solution of $\Pc^{(i)}$.
        \Step \textbf{3)} Set $\abf^{(i+1)}=\abf^{(i)} + \gamma^{(i)}(\hat{\abf}(\abf^{(i)})-\abf^{(i)})$.
        \Step \textbf{4)} $i\leftarrow i+1$ and go to step 1.
\end{algorithmic}
\end{algorithm}

\vspace{0.5em}
To briefly explain why SCA, specifically the NOVA algorithm works, consider the case of $\gamma^{(i)}=1$ in \eqref{eq:stepsizeTheo1}. Then, the current approximation point $\abf^{(i)}$ is the solution of the convex problem $\Pc^{(i-1)}$.
Note that the original problem $\Pc$ and the approximating convex problem $\Pc^{(j)}$ for any $j$ have the same convex objective function, and the feasible set of the approximating problem $\Pc^{(j)}$ for any $j$  is contained in the feasible set of the original problem $\Pc$  because $\tilde{c}_l(\abf,\bbf) \le 0$ implies $c_l(\abf)\le 0$ for any $\abf,\bbf$ due to the condition C4.
Hence,  the solution $\abf^{(i)}$ of $\Pc^{(i-1)}$ is a feasible point of the original problem $\Pc$, i.e., $c_l(\abf^{(i)})\le 0$. Now, consider the new approximating problem $\Pc^{(i)}$ around $\abf^{(i)}$ with constraints $\tilde{c}_l(\abf,\abf^{(i)}) \le 0$. By the condition C3 and the fact of $c_l(\abf^{(i)})\le 0$, we have
$\tilde{c}_l(\abf^{(i)},\abf^{(i)})= c_l(\abf^{(i)}) \le   0$. Hence, $\abf^{(i)}$ is a point in the feasible set of the new convex problem $\Pc^{(i)}$. Therefore, the solution $\abf^{(i+1)}$ of $\Pc^{(i)}$ is better (at least not worse) than  $\abf^{(i)}$ and monotone improvement  is achieved by iteration.

Now consider application of Theorem 1 to Problem \ref{problem:joint_beam}. Since the objective function \eqref{eq:prob1_obj} is linear and the constraints \eqref{eq:prob1_null}, \eqref{eq:prob1_norm} and \eqref{eq:prob1_power}  are convex, we need to convexify  the three non-convex constraints \eqref{eq:prob1_R1}, \eqref{eq:prob1_R21} and \eqref{eq:prob1_R22} to apply SCA in Theorem 1.    The key point of SCA is to obtain proper convex constraints approximating the original non-convex constraints, and this is the major step in SCA.
Although direct application of Theorem 1 is not easy due to
the complicated structure of \eqref{eq:prob1_R1}, \eqref{eq:prob1_R21} and \eqref{eq:prob1_R22},
Proposition \ref{prop:THP_NOMA_joint} shows that Problem \ref{problem:joint_beam} can be solved with SCA by introducing proper new slack variables, using the first-order Taylor expansion and applying some bounding technique  to approximate the original problem  to a convex optimization problem.

\vspace{0.5em}

\begin{proposition}\label{prop:THP_NOMA_joint}
Problem \ref{problem:joint_beam} can be solved by the proposed iterative algorithm based on SCA, presented in Appendix.  The proposed algorithm satisfies the conditions in Theorem 1, which guarantees convergence of the proposed algorithm.
\end{proposition}

\vspace{0.5em}

\textit{Proof)} Here, we briefly provide the sketch of  proof. See Appendix for the details.

For convex approximation for Problem 1, we need to convexify the three non-convex contraints \eqref{eq:prob1_R1}, \eqref{eq:prob1_R21} and \eqref{eq:prob1_R22}.  That is, we need to obtain an approximating convex constraint for each of these three non-convex constraints, but more importantly the obtained approximating convex constraints should satisfy the conditions C1 - C8 in Theorem 1
 in order to successfully apply  SCA  to Problem 1.
Here, we explain our main techniques for such approximation of the  non-convex constraint \eqref{eq:prob1_R1}, which is rewritten in the below for convenience:
\begin{equation} \label{eq:prob1_R1_proof}
|\hbf_{k1}^H\wbf_k|^2p_{k1} \ge \eta\frac{P}{N_c} |\Pibf_{\Hbf_1^{<k}}^\perp \hbf_{k1}|^2, ~\mbox{for each}~k.
\end{equation}
 The  approximation procedures for \eqref{eq:prob1_R21} and \eqref{eq:prob1_R22} are explained in Appendix.
 The convex approximation procedure for  \eqref{eq:prob1_R1_proof} is as follows.

\noindent $\bullet$ We first modify the constraint by introducing  slack variables to reduce its complexity. Note that the left-hand side (LHS) of \eqref{eq:prob1_R1_proof} is in the form of multiplication of two functions of  optimization variables $\wbf_k$ and $p_{k1}$, which is complicated to obtain an upper bound that satisfies the condition C4 in Theorem 1. Thus, we first relax this multiplication form by introducing slack variables of exponential forms\cite{Li&Chang&Lin&Chi:13SP} to exploit the fact that multiplication of exponential functions is  the exponential function of the linear sum of their arguments. We introduce $e^{l_{k1}}$ for $|\hbf_{k1}^H\wbf_k|^2$   and  $e^{m_{k1}}$ for $p_{k1}$. Then, we have
    \begin{align}
    & \eta\frac{P}{N_c}|\Pibf_{\Hbf_1^{<k}}^\perp \hbf_{k1}|^2 \le e^{l_{k1}+m_{k1}},\label{eq:proof1}\\
    &  e^{l_{k1}} \le p_{k1},\label{eq:proof2}\\
    &  e^{m_{k1}} \le |\hbf_{k1}^H\wbf_k|^2~(=\wbf_k^H (\hbf_{k1}\hbf_{k1}^H)\wbf_k).\label{eq:proof3}
    \end{align}
Note that the LHS of  \eqref{eq:prob1_R1_proof} is converted to a simple exponential function of the linear sum of two slack variables in \eqref{eq:proof1}. To compensate for this substitution,  eqs. \eqref{eq:proof2} and \eqref{eq:proof3} are newly added. The directions of the inequalities are determined to be consistent with the original constraint \eqref{eq:prob1_R1_proof}.
The resulting new three constraints \eqref{eq:proof1}, \eqref{eq:proof2} and \eqref{eq:proof3} implementing \eqref{eq:prob1_R1_proof} are in the form of (a convex function $\le$ a convex function).  \eqref{eq:proof2} is already a convex constraint since the larger side of \eqref{eq:proof2} is a linear function.

\noindent $\bullet$   Note that the desired constraint form in Theorem 1 is (a convex function $\le 0$) and the difference of two strictly convex functions as shown in \eqref{eq:proof1} and \eqref{eq:proof3}  is not convex.  Thus, we need to process \eqref{eq:proof1} and \eqref{eq:proof3} further.
  By obtaining a linear lower-bound for the larger side of each inequality, we can approximate the constraint to a convex constraint. For this, we use the first-order Taylor expansion.
 By applying the first-order Taylor expansion to the larger side of each of \eqref{eq:proof1} and \eqref{eq:proof3} at the current point $(\bar{l}_{k1},\bar{m}_{k1},\bar{\wbf}_{k})$, which corresponds to $\abf^{(i)}$ in Theorem 1,
\eqref{eq:proof1} and \eqref{eq:proof3} are approximated to two convex constraints satisfying C2 in Theorem 1:
    \begin{align}
    & \eta\frac{P}{N_c}|\Pibf_{\Hbf_1^{<k}}^\perp \hbf_{k1}|^2 \le e^{\bar{l}_{k1}+\bar{m}_{k1}}\left(1+l_{k1}+m_{k1}-\bar{l}_{k1}-\bar{m}_{k1}\right),\label{eq:proof1-1}\\
    &e^{m_{k1}} \le |\hbf_{k1}^H\bar{\wbf}_k|^2 + \langle\hbf_{k1}\hbf_{k1}^H\bar{\wbf}_k,\wbf_k-\bar{\wbf}_k\rangle.\label{eq:proof3-1}
\end{align}
where $\langle\abf,\bbf\rangle=2\mathrm{Re}(\abf^H\bbf)$. Note that for a given convex function, the first-order Taylor expansion is a linear lower bound of the convex function and it has the same function value and gradient value as the original function at the point of expansion.
Since the first-order Taylor expansion was applied to the larger side of the inequality,  the obtained approximating convex constraints satisfy the conditions C3, C4, C6 in Theorem 1.\footnote{That is, each  non-convex constraint is in the form of $c_{l,L}(\abf) \le c_{l,U}(\abf)$, i.e., $c_l(\abf):=c_{l,L}(\abf) - c_{l,U}(\abf) \le 0$ with $c_{l,L},c_{l,U}$ convex. Let $t_l(\abf,\abf^{(i)})$ be the first-order Taylor expansion of $c_{l,U}(\abf)$ at $\abf^{(i)}$. Then, $t_l(\abf,\abf^{(i)}) \le c_{l,U}(\abf)$ and $t_l(\abf^{(i)},\abf^{(i)}) = c_{l,U}(\abf^{(i)})$. Hence, we set the approximating convex constraint as  $\tilde{c}_l(\abf,\abf^{(i)}):= c_{l,L}(\abf) -t_l(\abf,\abf^{(i)})\le 0$, which satisfies C3,C4,C6 in Theorem 1.}
In addition, it is easy to check the validity of the continuity and differentiability  conditions C1, C5, C7 in Theorem 1.

\noindent $\bullet$
We can approximate  \eqref{eq:prob1_R21} and \eqref{eq:prob1_R22} with  convex constraints  by applying similar techniques, but  additional techniques are necessary to simplify \eqref{eq:prob1_R22} due to the  complexity of $I_k$ shown in \eqref{eq:THP_NOMA_interf33}. Basically, $I_k$ is too complicated for convex approximation. Although convex approximation with the exact $I_k$ is possible, it leads to a complicated convex problem with too many slack variables.
 Hence, we use an upper bound of $I_k$ and compute the lower bound of the second term in the minimum in  \eqref{eq:THP_NOMA_rate22} for $R_{k2}$. Then, we maximize this lower bound of  $R_{k2}$.
The details of convex approximations of \eqref{eq:prob1_R21} and \eqref{eq:prob1_R22} are in Appendix.

\noindent $\bullet$ The point used for the first-order Taylor expansion is the approximation point $\abf^{(i)}$ in Theorem 1. We can iteratively update the approximation point $\abf^{(i)}$  by using the solution of the approximating  convex optimization problem like the NOVA algorithm in Theorem 1. In this way, we obtain a sequence of solutions of the approximating convex problems, which converges to a stationary point of Problem \ref{problem:joint_beam} with  $R_{k2}$ replaced by the above-mentioned lower bound by Theorem 1, since the approximating constraints satisfy the conditions in Theorem 1.
\hfill $\blacksquare$

Problem \ref{problem:sequetial_beam} is a simpler version of Problem 1, and hence it can be solved in a similar way.

\section{Numerical Results}
\label{sec:numerical_results}

In this section, we provide some numerical results to evaluate the performance of the proposed user scheduling, beam design and power allocation algorithm for MU-MISO NOMA downlink. The basic setting for the simulations  in this section is as follows:  The AWGN variance was set to be one, i.e., $\sigma^2=1$ throughout the simulations. Each element of the channel vector for each strong user in $\Kc_1$ was generated independently from $\Cc\Nc(0,\sigma_{h,1}^2)$ with $\sigma_{h,1}^2=1$, whereas each element of the channel vector for each weak user in $\Kc_2$ was generated independently from $\Cc\Nc(0,\sigma_{h,2}^2)$ with $\sigma_{h,1}^2=0.01$. Hence, we have 20 dB difference in channel quality between the strong and weak users. ($|\Kc_1|=|\Kc_2|=K_{tot}/2$.)
\begin{figure}[!hbtp]
\centerline{\includegraphics[width=8cm]{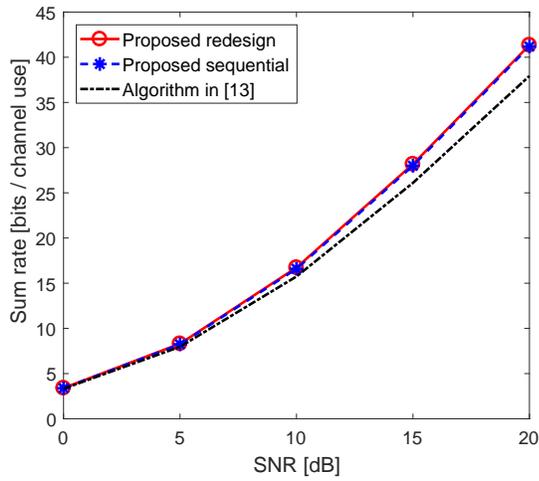}}
\centerline{(a)}
\centerline{\includegraphics[width=8cm]{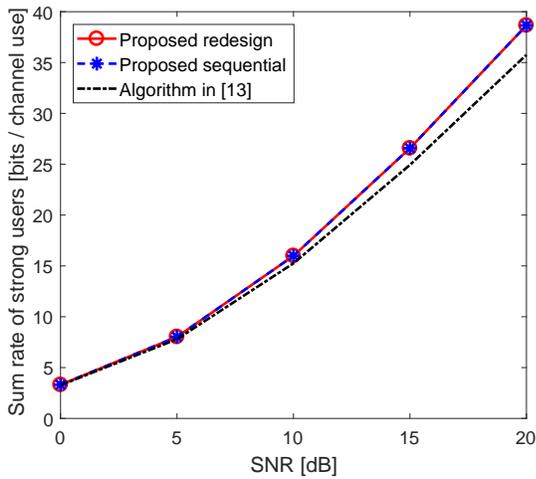}}
\centerline{(b)}
\centerline{\includegraphics[width=8cm]{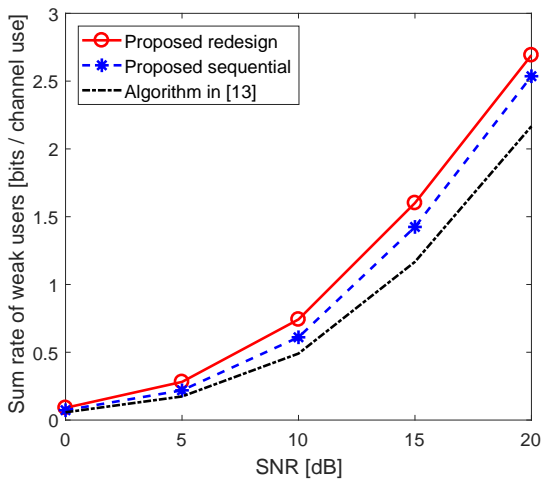}}
\centerline{(c)}
\caption{The average rate versus $10\log_{10}(P/\sigma^2)$) ($N_t=8$, $K_{tot}=200$): (a) total sum rate, (b) sum rate of strong users, and (c) sum rate of weak users } \label{fig:ave_sumrate_THP1}
\end{figure}

As the comparison baseline, we considered several hierarchical design methods for MU-MISO NOMA, the  design method in  \cite{Seo&Sung:18SP}, the NOMA-ZFBF-UMPS algorithm in \cite{Liu15:ICCW}, and the NOMA-FOUS algorithm in \cite{Sayed17:WD}. The design method in   \cite{Seo&Sung:18SP} is based on ZF inter-cluster beamforming and uses the two-user Pareto-optimal beam design and power allocation for the strong and weak users in each cluster forming a MISO broadcast channel with superposition coding and SIC. Thus, the intra-cluster design used in this method is optimal in the Pareto-optimality sense. For each cluster this method  sets a certain target SNR for the strong user and maximizes the weak user rate. Hence, this method  has the capability of trading off the strong user rate for the weak user rate by changing the target SNR for the strong user.
It is shown in  \cite{Seo&Sung:18SP} that this method outperforms several other user scheduling and power allocation method based on ZF inter-cluster beamforming.  The original algorithms in \cite{Sayed17:WD}, \cite{Liu15:ICCW} consider only a single set of users for selection of both strong and weak users. Hence, the simulation setting in \cite{Sayed17:WD}, \cite{Liu15:ICCW}  is different. So, we modified the original two algorithms to make the strong user be selected from $\Kc_1$ and the weak user be selected from $\Kc_2$.

Fig. \ref{fig:ave_sumrate_THP1} shows the performance of the proposed THP-aided algorithm as compared to the ZF intercluster-beamforming-based method in  \cite{Seo&Sung:18SP}. The figure shows the average rate versus SNR defined as $10\log_{10}\frac{P}{\sigma^2}$ for $N_t=8$, $K_{tot}=200$ (i.e.,
 $|\Kc_1|=|\Kc_2|=100$).
The value of the strong user target SNR parameter $\eta$ was set as $\eta=0.3$ for the proposed method, whereas $\eta=0.4$ (this second $\eta$ as defined in \cite{Seo&Sung:18SP}) for the ZF intercluster-beamforming-based method in  \cite{Seo&Sung:18SP}. (The definition of $\eta$  in \cite{Seo&Sung:18SP} is a bit different from that in this paper, but the role is the same.)
The average rates were obtained by averaging $50$ independent channel realizations.  It is seen that the proposed THP-aided method outperforms the  ZF intercluster beamforming-based method  in  \cite{Seo&Sung:18SP}. It is also seen that the beam redesign and power allocation by solving Problem 1 yields non-trivial gain over the initial sequential greedy beam design with equal cluster power allocation in Section \ref{subsec:user_scheduling}.

\begin{figure}[t]
\centering
\includegraphics[width=9cm]{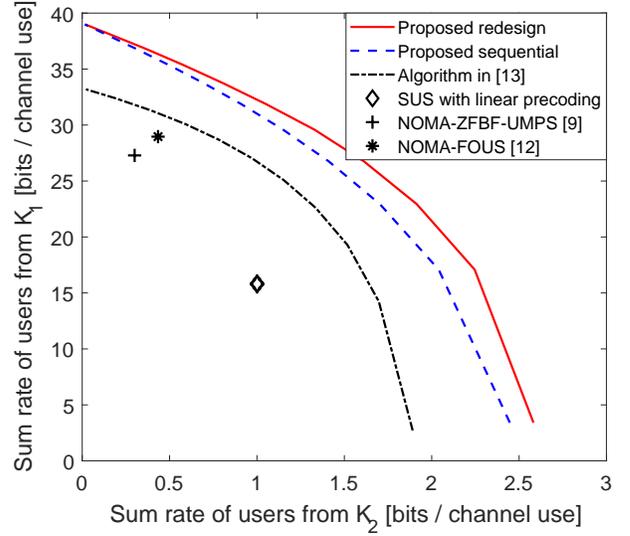}
\caption{ Sum strong user rate versus sum weak user rate with sweeping $\eta$ ($N_t=8$, $|\Kc_1|=|\Kc_2|=100$, and $10\log_{10}\frac{P}{\sigma^2}=15$dB)}
\label{fig:rateRegion_THP}
\end{figure}

Fig. \ref{fig:rateRegion_THP} shows the sum strong user rate versus the sum weak user rate by sweeping the strong user target SNR parameter $\eta$ in Problem 1 for $N_t=8$, $|\Kc_1|=|\Kc_2|=100$, and $10\log_{10}\frac{P}{\sigma^2}=15$ dB. Note that the algorithm in \cite{Seo&Sung:18SP} has the capability of trading off the strong user rate for the weak user rate, whereas  the NOMA-ZFBF-UMPS algorithm and the NOMA-FOUS algorithm do not. It is seen that the proposed THP-based method enlarges the rate region noticeably as compared to the existing ZF intercluster beamforming-based methods.  Again, it is  seen that the beam redesign and power allocation by solving Problem 1 yields non-trivial gain over the initial sequential greedy beam design with equal cluster power allocation in Section \ref{subsec:user_scheduling}.

\section{Conclusion}
\label{sec:conclusion}

In this paper, we have considered THP for MU-MISO NOMA donwlink systems. We have applied THP   under  the hierarchical structure in which multiple clusters each with two users are formed and served in the spatial domain and users in each cluster are served in the power domain. The application of THP  eliminates  ICI  to the strong users and enlarges the dimension of the beam design space, which can be exploited for inter-cluster beam design to mitigate ICI to weak users on top of weak user selection.
Exploiting this enlarged beam design space, we have proposed a two-step user scheduling algorithm together with two beam design methods: sequential greedy beam design and after-user-selection beam redesign and power allocation.
To solve the design problems, we have proposed an efficient algorithm based on SCA, which guarantees convergence to a stationary point of the problem. Numerical results show that the proposed THP-aided beam design and user scheduling yield noticeable gain over existing ZF inter-cluster beamforming-based methods. Furthermore, the introduced two-step technique for convexification for SCA can be useful to other general problems requiring convex approximation.

\section*{Appendix}
\label{sec:append}

\textit{Proof of Proposition \ref{prop:THP_NOMA_joint}:}
Since in Problem \ref{problem:joint_beam} the objective function \eqref{eq:prob1_obj} is linear and the constraints \eqref{eq:prob1_null}, \eqref{eq:prob1_norm} and \eqref{eq:prob1_power}  are convex, we need to obtain approximating convex constraints for  the three non-convex constraints \eqref{eq:prob1_R1}, \eqref{eq:prob1_R21} and \eqref{eq:prob1_R22}   satisfying  the conditions C1 - C8 in  Theorem 1. Then, we can apply the NOVA algorithm to obtain a stationary point of Problem 1.

The non-convex constraints \eqref{eq:prob1_R1}, \eqref{eq:prob1_R21} and \eqref{eq:prob1_R22} of Problem \ref{problem:joint_beam} are rewritten here for convenience as
\begin{align}
|\hbf_{k1}^H\wbf_k|^2p_{k1}&\ge \eta\frac{P}{N_c} |\Pibf_{\Hbf_{k}}^\perp \hbf_{k1}|^2,  \label{eq:append_cond1}\\
R_{k2} &\le \log_2\left(1+\frac{|\hbf_{k1}^H\wbf_k|^2p_{k2}}{|\hbf_{k1}^H\wbf_k|^2p_{k1}+\sigma^2}\right),  \label{eq:append_cond2}\\
R_{k2} &\le \log_2\left(1+\frac{|\hbf_{k2}^H\wbf_k|^2p_{k2}}{{I}_k + \sigma^2}\right), \label{eq:append_cond3}
\end{align}
where
\begin{align}
I_k &= |\hbf_{k2}^H\wbf_k|^2p_{k1} + \sum_{j<k}\left|\hbf_{k2}^H\wbf_j- \frac{\hbf_{k2}^H\wbf_k}{\hbf_{k1}^H\wbf_k}
\hbf_{k1}^H\wbf_j
\right|^2 p_j\nonumber\\
& ~~~~~~~~~~~~~~~~~~~+ \sum_{j>k}|\hbf_{k2}^H{\wbf_j}|^2p_j.\label{eq:append_interf_term}
\end{align}

{\it Step 1)} First, we convert the three non-convex constraints into the form of (a convex function $\le$ a convex function) by introducing slack variables $\bigcup_{k=1}^{N_c} \{{m}_{k1},{m}_{k2},{m}_{k3},{l}_{k1},{l}_{k2},{l}_{k3},{l}_{k4},
{n}_{kj} ~\forall j\ne k\}$ as follows:

{i)} $|\hbf_{k1}^H\wbf_k|^2p_{k1}\ge \eta\frac{P}{N_c} |\Pibf_{\Hbf_{k}}^\perp \hbf_{k1}|^2$ (eq.\eqref{eq:append_cond1}) :
\begin{align}
&e^{l_{k1}+m_{k1}}\ge\eta\frac{P}{N_c}|\Pibf_{\Hbf_{k}}^\perp \hbf_{k1}|^2\label{eq:first1},\\
&e^{l_{k1}}\le p_{k1}\label{eq:first2},\\
&e^{m_{k1}}\le \left|\hbf_{k1}^H\wbf_k\right|^2\label{eq:first3},
\end{align}

{ii)} $R_{k2} \le \log_2\left(1+\frac{|\hbf_{k1}^H\wbf_k|^2p_{k2}}{|\hbf_{k1}^H\wbf_k|^2p_{k1}+\sigma^2}\right)$ (eq. \eqref{eq:append_cond2}):
\begin{align}
&\left(2^{R_{k2}}-1\right)\cdot \left(\sigma^2+e^{l_{k2}+m_{k1}}\right) \le e^{l_{k3}+m_{k1}}\label{eq:second1},\\
&e^{l_{k2}}\ge p_{k1}\label{eq:second2},\\
&e^{l_{k3}}\le p_{k2}\label{eq:second3},
\end{align}
where \eqref{eq:second1} can further be changed in the form of (a convex function $\le$ a convex function) as
\begin{equation}
\sigma^2 2^{R_{k2}}+ 2^{R_{k2}}e^{l_{k2}+m_{k1}} \le \sigma^2+e^{l_{k2}+m_{k1}}+e^{l_{k3}+m_{k1}}.
\end{equation}
The procedure for {\em i)} is already explained in the below of Proposition 1, and  a similar procedure is applied to {\em ii)}.

{iii)} $R_{k2} \le \log_2\left(1+\frac{|\hbf_{k2}^H\wbf_k|^2p_{k2}}{{I}_k + \sigma^2}\right)$ (eq. \eqref{eq:append_cond3}): Convex approximation of this constraint is  complicated due to the structure of $I_k$ as shown in \eqref{eq:append_interf_term}.
The first and third terms in the RHS of \eqref{eq:append_interf_term} are in the form of the product of a power term and a quadratic term of $\wbf_k$, but  the second term of the RHS of \eqref{eq:append_interf_term} is not simple.
 Although convex approximation with the exact $I_k$ is possible, this leads to a complicated convex problem with too many slack variables.
 Hence, we use an upper bound of the second term of the RHS of \eqref{eq:append_interf_term} to simplify the problem.
 The second term of the RHS of \eqref{eq:append_interf_term}
  can be upper bounded as
\begin{align}
&\left|\hbf_{k2}^H\wbf_j-  \frac{\hbf_{k2}^H\wbf_k}{\hbf_{k1}^H\wbf_k}  {\hbf_{k1}^H\wbf_j}\right|^2p_j \nonumber\\
=& \frac{|\hbf_{k1}^H\wbf_k\cdot\hbf_{k2}^H\wbf_j-\hbf_{k2}^H\wbf_k\cdot\hbf_{k1}^H\wbf_j|^2p_j}{|\hbf_{k1}^H\wbf_k|^2} \\
\stackrel{(a)}{\le}&\frac{(|\hbf_{k1}^H\wbf_k|^2+|\hbf_{k2}^H\wbf_k|^2)\cdot(|\hbf_{k1}^H\wbf_j|^2+|\hbf_{k2}^H\wbf_j|^2)p_j}{|\hbf_{k1}^H\wbf_k|^2}\\
=& \frac{\wbf_k^H(\hbf_{k1}\hbf_{k1}^H + \hbf_{k2}\hbf_{k2}^H)\wbf_k\cdot\wbf_j^H(\hbf_{k1}\hbf_{k1}^H + \hbf_{k2}\hbf_{k2}^H)\wbf_jp_j}{|\hbf_{k1}^H\wbf_k|^2}\label{eq:upperbound}
\end{align}
where the Cauchy-Schwartz inequality is applied to step (a). By summing the term in \eqref{eq:upperbound} and the first and third terms in the RHS of \eqref{eq:append_interf_term}, we have an upper bound, denoted by $\bar{I}_k$, on $I_k$.  Substituting $\bar{I}_k$ into $I_k$ yields a lower bound on $R_{k2}$. Hence, by approximating \eqref{eq:append_cond3} with $\bar{I}_k$ instead of $I_k$, we modify the original problem, Problem 1, slightly as the problem of maximizing the lower bound of $\sum_k R_{k2}$. (Note that a lower bound is taken only on the second term of the RHS of \eqref{eq:THP_NOMA_rate22}.)
Since $\Pibf_k:=\hbf_{k1}\hbf_{k1}^H + \hbf_{k2}\hbf_{k2}^H$  in the numerator of \eqref{eq:upperbound} is a positive-definite matrix, we can substitute the terms $p_j$, $\wbf_k^H\Pibf_k\wbf_k$, $\wbf_j^H\Pibf_k\wbf_j$ and $|\hbf_{k1}^H\wbf_k|^2$ in \eqref{eq:upperbound} by exponential functions with new slack variables. Then, the constraint \eqref{eq:append_cond3} with $I_k$ replaced by $\bar{I}_k$ is rewritten as
\begin{align}
&(2^{R_{k2}}-1)\cdot(\tilde{I}_k+\sigma^2) \le e^{l_{k3}+m_{k2}},\\
&e^{m_{k2}} \le |\hbf_{k2}^H\wbf_k|^2,\\
&e^{m_{k3}} \ge \wbf_k^H(\hbf_{k1}\hbf_{k1}^H + \hbf_{k2}\hbf_{k2}^H)\wbf_k,\\
&e^{l_{k4}} \ge p_{k} ~~~(\mbox{i.e.,~}e^{l_{j4}} \ge p_{j}) ,\\
&e^{n_{kj}} \ge \left\{\begin{array}{ll}\wbf_j^H(\hbf_{k1}\hbf_{k1}^H + \hbf_{k2}\hbf_{k2}^H)\wbf_j,&\mbox{for }j< k,\\
 |\hbf_{k2}^H\wbf_j|^2,&\mbox{for }j> k
\end{array}\right.
\end{align}
where
\begin{align}
\tilde{I}_k &= e^{l_{k2}+m_{k2}} + e^{m_{k3}-m_{k1}}\sum_{j<k}e^{l_{j4}+n_{kj}} + \sum_{j>k} e^{l_{j4}+n_{kj}}.
\end{align}
Note that the directions of the inequalities with the newly introduced slack variables are determined  to maintain consistency with the original non-convex optimization.

{\it Step 2)} In Step 1, by introducing the slack variables, we expressed the three non-convex constraints as multiple inequalities each in the form of  $c_{l,L}(\abf) \le c_{l,U}(\abf)$, i.e., $c_l(\abf):=c_{l,L}(\abf) - c_{l,U}(\abf) \le 0$,  with $c_{l,L},c_{l,U}$ convex in the optimization variables $\abf$. To obtain the desired form of the approximating convex constraint described in Theorem 1,
we apply the first-order Taylor expansion to the larger side of each of the inequality constraints obtained in Step 1 at the current point of Taylor expansion, which corresponds to $\abf^{(i)}$ in Theorem 1.
That is, let $t_l(\abf,\abf^{(i)})$ be the first-order Taylor expansion of $c_{l,U}(\abf)$ at $\abf^{(i)}$. Then, $t_l(\abf,\abf^{(i)}) \le c_{l,U}(\abf)$ and $t_l(\abf^{(i)},\abf^{(i)}) = c_{l,U}(\abf^{(i)})$ since the first-order Taylor expansion for a given convex function is a linear lower bound of the convex function and  has the same function value  as the original function at the point of expansion. Furthermore, it has the same gradient value as the original function at the point of expansion. Hence, we obtain a desired approximating convex constraint as $\tilde{c}_l(\abf,\abf^{(i)}):= c_{l,L}(\abf) -t_l(\abf,\abf^{(i)})\le 0$ for each  $c_l(\abf)=c_{l,L}(\abf) - c_{l,U}(\abf) \le 0$. Then,
  the approximating convex constraints  satisfy C3, C4, C6 as well as
 the easily-verifiable continuity and differentiability  conditions C1, C5, C7 in Theorem 1.
 Thus, we finally obtain a convex optimization problem approximating Problem 1 with $I_k$ replaced by $\bar{I}_k$, given by $\Pc_{1}(\bar{\alphabf})$ in the next page.
In $\Pc_{1}(\bar{\alphabf})$,
\[
\bar{\alphabf}:= [\{\bar{\wbf}_k, \bar{m}_{k1},\bar{m}_{k2},\bar{m}_{k3},\bar{l}_{k1},\bar{l}_{k2},\bar{l}_{k3},\bar{l}_{k4},\bar{n}_{kj}~ \forall j\ne k\}_{k=1}^{N_c}]
\]
is the point at which the first-order Taylor series is obtained, and
\begin{align}
{f}(x,\bar{x}) &:=e^{\bar{x}}(1+x-\bar{x}),\\
{g}_{\cbf}(\dbf,\bar{\dbf}) &:= \left|\cbf^H\bar{\dbf}\right|^2+\langle\cbf\cbf^H\bar{\dbf},\dbf-\bar{\dbf}\rangle,
\end{align}
where $\langle\cbf,\dbf\rangle = 2\mathrm{Re}\left(\cbf^H\dbf\right)$.  (These two functions are  the first-order Taylor series of $\tilde{f}(x):=e^{x}$ at $\bar{x}$ and $\tilde{g}_\cbf(\dbf):=|\cbf^H\dbf|^2$ at $\bar{\dbf}$, respectively.)

In $\Pc_{1}(\bar{\alphabf})$,    \eqref{eq:P1app1}-\eqref{eq:P1app2} correspond to \eqref{eq:append_cond1};  \eqref{eq:P1app3}-\eqref{eq:P1app4} correspond to \eqref{eq:append_cond2}; and \eqref{eq:P1app5}-\eqref{eq:P1app6} and \eqref{eq:P1app2} with $q=2$ correspond to
\eqref{eq:append_cond3}.   Since $\Pc_{1}(\bar{\alphabf})$ is a convex problem, it can be solved by any convex optimization solver.

{\it Step 3)}
Now, we propose an algorithm that iteratively solves $\Pc_1(\bar{\alphabf})$ by updating its parameter vector $\bar{\alphabf}$. Let $\bar{\alphabf}^{(i)}$ be the Taylor expansion parameter vector at iteration $i$, given by
\begin{align}
\bar{\alphabf}^{(i)}:= [\{\bar{\wbf}_k^{(i)}, &\bar{m}_{k1}^{(i)},\bar{m}_{k2}^{(i)},\bar{m}_{k3}^{(i)},\bar{l}_{k1}^{(i)},\bar{l}_{k2}^{(i)},\bar{l}_{k3}^{(i)},\bar{l}_{k4}^{(i)},\nonumber\\
&~~~~~~~~~~~~~~~~~~~~~~~~~\bar{n}_{kj}^{(i)}~ \forall j\ne k\}_{k=1}^{N_c}].
\end{align}

\begin{algorithm}[!b]
   \caption{\textbf{{Joint Beam Design and Power Allocation}}}\label{THP:algorithm2}
    \begin{algorithmic}[1]
       \State \textbf{Initialization}:
       \State Initialize $\bar{\alphabf}^{(0)}$.
       \State Set the stopping parameter $\epsilon$.
       \State $i\leftarrow0$, $R_{k2}^{(-1)}\leftarrow 0$, $R_{k2}^{(-2)}\leftarrow0$, $\forall k$.
       \While{$i=0$ or $(\sum_kR_{k2}^{(i-1)}-\sum_kR_{k2}^{(i-2)})<\epsilon$}
       \State Solve  $\Pc_1(\bar{\alphabf}^{(i)})$ and obtain its solution  $\Sc^{(i)}$.
       \State Update $\bar{\alphabf}^{(i+1)}$ with $\Sc^{(i)}$.
       \State $i\leftarrow i+1$
       \EndWhile
\end{algorithmic}
\end{algorithm}

The proposed algorithm is basically an application of the NOMA algorithm \cite{Scutari&Facchinei&Lampariello:17SP} shown in Algorithm 2.  The proposed algorithm is summarized in Algorithm 3. In line 7 of Algorithm 3, for the update of the Taylor expansion point, we used the setting that corresponds to $\gamma^{(i)}=1$ in Step 3 of  the NOVA algorithm shown in Algorithm 2.
We initialize $\bar{\alphabf}^{(0)}$ as follows:  First, with the  convexification techniques in Steps 1 and 2, we solve the simpler sequential problem, Problem 2 with initialization of $p_{k1}^{(0)}$, $p_{k2}^{(0)}$, and $\wbf_k^{(0)}$ as in lines 3 - 8 in Algorithm 1 together with slack variable initialization:
\begin{align} \label{eq:T_update}
\bar{m}_{kq}^{(0)} &= \log(|\hbf_{kq}^H\wbf_k^{(0)}|^2),~\mbox{for }q=1,2,\nonumber\\
\bar{m}_{k3}^{(0)} &= \log((\wbf_k^{(0)})^H(\hbf_{k1}\hbf_{k1}^H+\hbf_{k2}\hbf_{k2}^H)\wbf_k^{(0)}),\nonumber\\
\bar{l}_{k1}^{(0)} &= \bar{l}_{k2}^{(0)} = \log(p_{k1}^{(0)}),\nonumber\\
\bar{l}_{k3}^{(0)} &= \log(p_{k2}^{(0)}),\nonumber\\
\bar{l}_{k4}^{(0)} &= \log(p_{k}^{(0)}),\nonumber\\
\bar{n}_{kj}^{(0)} &= \left\{
\begin{array}{ll}\log((\wbf_j^{(0)})^H(\hbf_{k1}\hbf_{k1}^H+\hbf_{k2}\hbf_{k2}^H)\wbf_j^{(0)}),&\forall j<k\\
\log(|\hbf_{k2}^H\wbf_j^{(0)}|^2)&\forall j>k.\end{array}\right.
\end{align}
Then, with the obtained solution from Problem 2, we initialize $p_{k1}^{(0)}$, $p_{k2}^{(0)}$, and $\wbf_k^{(0)}$ for Problem 1 and slack variables for Problem 1 with the same way as in \eqref{eq:T_update}.

\begin{figure*}
\begin{align}
&\Pc_{1}(\bar{\alphabf}):\mbox{The Approximated Convex Problem} \nonumber\\
&\mathop{\max}\limits_{R_{k2},\wbf_k,p_{k1},p_{k2},\forall k, \mbox{and slack variabls}} ~~\sum_k  R_{k2}\\
&~~~~~~~~~~~~~~~~~~\mbox{subject to }\nonumber\\
& f(l_{k1}+m_{k1},\bar{l}_{k1}+\bar{m}_{k1})\ge\eta\frac{P}{N_c}|\Pibf_{\Hbf_{k}}^\perp \hbf_{k1}|^2,~\forall k \label{eq:P1app1}\\
& e^{l_{k1}} \le p_{k1},~\forall k\label{eq:slack1}\\
& e^{m_{kq}}\le g_{\hbf_{kq}}(\wbf_k,\bar{\wbf}_k),~\forall k,~q=1,2 \label{eq:P1app2}\\
& \nonumber\\
&e^{R_{k2}\ln 2+l_{k2}+m_{k1}}+\sigma^2e^{R_{k2}\ln 2}\le \sigma^2
+  f(l_{k2}+m_{k1},\bar{l}_{k2}+\bar{m}_{k1})+ f(l_{k3}+m_{k1},\bar{l}_{k3}+\bar{m}_{k1}),~\forall k \label{eq:P1app3}\\
& f(l_{k2},\bar{l}_{k2})\ge p_{k1},~\forall k\\
& e^{l_{k3}}\le p_{k2},~\forall k \label{eq:P1app4}\\
&\nonumber\\
& e^{R_{k2}\ln 2}(\tilde{I}_k+\sigma^2)\le \sigma^2 + f(l_{k2}+m_{k2},\bar{l}_{k2}+\bar{m}_{k2})+\sum_{j<k}f(m_{k3}-m_{k1}+l_{j4}+n_{kj},\bar{m}_{k3}-\bar{m}_{k1}+\bar{l}_{j4}+\bar{n}_{kj})\nonumber\\
&~~~~~~~~~~~~~~~~~~~~~~~~~~~~~~~~~~~~~~~~~~~~~~~~~~~~~~~~~~~~+\sum_{j>k}f(l_{j4}+n_{kj},\bar{l}_{j4}+\bar{n}_{kj})+f(l_{k3}+m_{k2},\bar{l}_{k3}+\bar{m}_{k2}),~\forall k \label{eq:P1app5}\\
& f(m_{k3},\bar{m}_{k3})\ge \wbf_k^H(\hbf_{k1}\hbf_{k1}^H + \hbf_{k2}\hbf_{k2}^H)\wbf_k,~\forall k\\
& f(l_{k4},\bar{l}_{k4})\ge p_k,~\forall k\\
& f(n_{kj},\bar{n}_{kj})\ge\left\{\begin{array}{ll}\wbf_j^H(\hbf_{k1}\hbf_{k1}^H + \hbf_{k2}\hbf_{k2}^H)\wbf_j,&\forall j< k\\
|\hbf_{k2}^H\wbf_j|^2,&\forall j> k
\end{array}\right. , \forall k \label{eq:P1app6}\\
&\nonumber\\
& (\Hbf_{1}^{<k})^H\wbf_k = \mathbf{0}, ~\forall k\\
& \|\wbf_k\| \le 1,~\forall k\\
& \sum_{k}(p_{k1} + p_{k2}) \le P.
\end{align}
\end{figure*}

{\it Step 4)} Finally, we prove that Algorithm \ref{THP:algorithm2} converges to a stationary point of Problem \ref{problem:joint_beam} with $I_k$ replaced by $\bar{I}_k$. It is already mentioned in Step 2 that the obtained approximating convex problem satisfies the conditions C1-C7 of Theorem 1. The final technical condition in Theorem 1 is Slater's condition C8.
Note that Slater's condition  requires that there exists an interior feasible point, i.e., a feasible point that satisfies every inequality constraint of the problem with strict inequality. We can state that if $\Pc_1(\bar{\alphabf}^{(i)})$ has a non-trivial solution such that $\sum_k R_{k2} \ne 0$, then there exists an interior feasible point for $\Pc_1(\bar{\alphabf}^{(i)})$.
Consider the solution of $\Pc_1(\bar{\alphabf}^{(i)})$, $\Sc^{(i)}=\{R_{k2}^{(i)\star},\wbf^{(i)\star},p_{k1}^{(i)\star},p_{k2}^{(i)\star},\forall k\mbox{ and slack variables}^\star\}$. By setting $R_{k2}^{(i)\star}=0,\forall k$ and fixing other variables in $\Sc^{(i)}$,  the constraints \eqref{eq:P1app3} and \eqref{eq:P1app5} involving $R_{k2}$ are satisfied with strict inequality. By exploiting the gap between the LHS and RHS of \eqref{eq:P1app3} and \eqref{eq:P1app5}, we can adjust other variables in $\Sc^{(i)}$ to have strict inequality for all other constraints. That is, since $R_{k2}^{(i)\star}$ is decreased, every slack variable related to interference such as $l_{k2}$, $l_{k4}$, $m_{k3}$ and $n_{kj}$ for $j\ne k$, can be increased while strict inequality for \eqref{eq:P1app3} and \eqref{eq:P1app5} is  remained. Then, the constraints inserted by introducing these slack variables satisfy strict inequality. Other slack variables can be adjusted with sufficiently small amount to have strict inequality for the remaining constraints.
 Thus, there exists a set of variables that satisfy all the constraints with strict inequality if $\Pc_1(\bar{\alphabf}^{(i)})$ has a non-trivial solution such that $\sum_k R_{k2} \ne 0$.

Therefore, by Theorem 1,  the proposed algorithm converges to a stationary point of Problem 1 with $I_k$ replaced by $\bar{I}_k$.
\hfill{$\blacksquare$}



\ifCLASSOPTIONcaptionsoff
  \newpage
\fi

\end{document}

%% file: macros.tex
\def\psfancypar#1#2{\begingroup\def\par{\endgraf\endgroup\lineskiplimit=0pt}
               \setbox2=\hbox{\large\sc #2}
               \newdimen\tmpht \tmpht \ht2 \advance\tmpht by \baselineskip
               \font\hhuge=Times-Bold at \tmpht
               \setbox1=\hbox{{\hhuge #1}}
               \count7=\tmpht \count8=\ht1
               \divide\count8 by 1000 \divide\count7 by \count8 
               \tmpht=.001\tmpht\multiply\tmpht by \count7 
               \font\hhuge=Times-Bold at \tmpht
               \setbox1=\hbox{{\hhuge #1}}
               \noindent
                \hangindent1.05\wd1
               \hangafter=-2 {\hskip-\hangindent
               \lower1\ht1\hbox{\raise1.0\ht2\copy1}%
                \kern-0\wd1}\copy2\lineskiplimit=-1000pt}

\newcommand{\E}{\mbox{{\rm E}}}

\newcommand{\abf}{\mbox{${\bf a}$}}

\def\boxit#1{\vbox{\hrule\hbox{\vrule\kern3pt
        \vbox{\kern3pt#1\kern3pt}\kern3pt\vrule}\hrule}}

\def\reals{ { {\rm  I \kern-0.15em R }  } }
\def\complex{ {\,{{\rm C} \kern-0.50em \raise0.20ex {  |}}\, }}
\def\alphabf{\hbox{\boldmath$\alpha$\unboldmath}}

\def\mubf{\hbox{\boldmath$\mu$\unboldmath}}

\def\Sigmabf{\hbox{$\bf \Sigma$}}

\def\Pibf{{\bf \Pi}}
\def\abf{{\bf a}}
\def\bbf{{\bf b}}
\def\cbf{{\bf c}}
\def\dbf{{\bf d}}

\def\gbf{{\bf g}}
\def\hbf{{\bf h}}

\def\nbf{{\bf n}}

\def\qbf{{\bf q}}

\def\sbf{{\bf s}}

\def\wbf{{\bf w}}
\def\xbf{{\bf x}}
\def\ybf{{\bf y}}

\def\xbf{{\bf x}}
\def\ybf{{\bf y}}
\def\Abf{{\bf A}}

\def\Hbf{{\bf H}}
\def\Ibf{{\bf I}}

\def\Lbf{{\bf L}}

\def\Qbf{{\bf Q}}
\def\Rbf{{\bf R}}

\def\Wbf{{\bf W}}

\def\Ac{{\cal A}}

\def\Cc{{\cal C}}

\def\Kc{{\cal K}}

\def\Nc{{\cal N}}

\def\Pc{{\cal P}}

\def\Sc{{\cal S}}

\def\Uc{{\cal U}}

\def\be{\vskip .3cm \begin{equation}}
\def\ee{\end{equation} \vskip .4cm \noindent}
\newcommand{\R}{\mbox{$\hat {\bf R}_{N}$}}

\def\Rxx{\Rbf_{\ssstyle X\kern-.1em X}}

\let\ssstyle=\scriptscriptstyle

\def\Kout{\setbox1=\hbox{\Huge\bf K}\hbox to
1.05\wd1{\hspace{.05\wd1}
}}
\def\Sout{\setbox1=\hbox{\Huge\bf S}\hbox to 1.05\wd1{\hspace{.05\wd1}
}}

%% file: labelfig.tex
  \ifx\LabelFigloaded\MYundefined\relax
  \else
   \fi

  \def\LabelFigloaded{\relax}

  \chardef\LabelFigCatAt\the\catcode`\@
  \catcode`\@=11

 \let\LabelFigwlog@ld\wlog
 \def\wlog#1{\relax}

 \ifx\\\MYundefined@
    \let\\\relax
 \fi

  \def\ms@g{\immediate\write16}

 \def\N@wif{\csname newif\endcsname }
 \def\Temp@ {\N@wif\ifIN@}
 \ifx\INN@\MYundefined@
    \else \let\Temp@\relax
 \fi
 \Temp@

  \def\IN@{\expandafter\INN@\expandafter}
  \long\def\INN@0#1@#2@{\long\def\NI@##1#1##2##3\ENDNI@
    {\ifx\m@rker##2\IN@false\else\IN@true\fi}%
     \expandafter\NI@#2@@#1\m@rker\ENDNI@}
  \def\m@rker{\m@@rker}
 
  \newtoks\Initialtoks@  \newtoks\Terminaltoks@
  \def\SPLIT@{\expandafter\SPLITT@\expandafter}
  \def\SPLITT@0#1@#2@{\def\TTILPS@##1#1##2@{%
     \Initialtoks@{##1}\Terminaltoks@{##2}}\expandafter\TTILPS@#2@}

 \def\Shifted@@#1#2#3{\setbox0=\hbox{#3}%
   \raise -\dp0\vbox {\kern-#2%
       \hbox {\kern#1\unhbox0\kern-#1}%
           \kern#2}}

 \newcount\gridcount
 \newbox\auxGridbox@ \newbox\hGridbox@ \newbox\vGridbox@
 \newbox\Labelbox@ \newbox\auxLabelbox@
 \newbox\Coordinatebox@
 \newtoks\Labeltoks@
 \newdimen\Wdd@ \newdimen\Htt@
 \newdimen\Wddd@ \newdimen\Httt@
 
 \def\Wr@{\immediate\write16}

 \newdimen\GL@wd
 \def\GridLineWidth#1{\GL@wd=#1}

 \def\gobble#1{}
 \def\EdgeErr@{\Wr@{}%
      \Wr@{\string\Edges\space argument
      1, 10, 100 or 1000 please\string!}%
      }

 \newcount\Edgect@

 \def\Sweepup#1\endSweepup{}

 \def\SetEdges@{%
    \edef\Zr@@s{\expandafter\gobble\number\Edgect@\empty}%
        \count255=0\Zr@@s\relax
        \ifnum\count255=\z@\else\EdgeErr@\show\tailtest\fi
        \count255=1\Zr@@s\relax
        \ifnum\count255=\Edgect@\relax\else\EdgeErr@\show\leadtest\fi
    \EdgGl@b\edef\Zr@s{\expandafter\gobble\Zr@@s\empty}
    \ifnum\Edgect@>\@ne\relax\EdgGl@b\let\L@Dc\empty
        \else\EdgGl@b\edef\L@Dc{\string.}\fi
    \ifnum\Edgect@>\@ne\relax
        \EdgGl@b\edef\Edgescale@##1{\divide##1 by \Edgect@}%
        \else\EdgGl@b\edef\Edgescale@##1{}\fi
    }

 \def\Edges#1{\Edgect@=#1\relax
     \let\EdgGl@b\global \SetEdges@}

 \Edges{1}

 \def\hhrule{\hrule height \GL@wd\vskip-.\GL@wd}

 \def\hRule@{%
   \advance\gridcount -2%
   \vfil\hhrule\vfil
   \llap{\smash{\raise -2.5pt
     \hbox{\L@Dc\number\gridcount\Zr@s\kern2pt}}}%
   \hhrule
   }

\def\vvrule{\vrule width \GL@wd \kern-\GL@wd}

 \def\vRule@{\advance\gridcount 2%
   \hfil\vvrule\hfil
   \setbox\auxGridbox@=\vbox to 0pt
      {\vskip \Htt@\vskip 2pt
        \hbox to 0pt{\hss\L@Dc\number\gridcount\Zr@s\hss}\vss}%
      \wd\auxGridbox@=0pt \box\auxGridbox@
   \vvrule
   }

 \def\PlaceGrid@@{\gridcount=10 
  \setbox\hGridbox@=\hbox{%
        \hbox{%
             \hskip-.4pt\vrule
             \vbox to \Htt@{%
               \offinterlineskip\parindent=\z@\relax
               \hbox to \Wdd@{\hfil}
               \hRule@\hRule@\hRule@\hRule@
               \vfil\hhrule\vfil}%
             \vrule\hskip-.4pt}
    }%
  \gridcount=0%
  \setbox\vGridbox@=\hbox{%
      \vbox{\offinterlineskip\parindent=0pt\hsize=0pt
         \vskip-.4pt\hrule%
         \hbox to \Wdd@{%
                 \vtop to \Htt@{\vfil}%
                 \vRule@\vRule@\vRule@\vRule@
                 \hfil\vvrule\hfil}%
         \hrule\vskip-.4pt}}%
  \wd\hGridbox@=0pt\ht\hGridbox@=0pt
  \wd\vGridbox@=0pt\ht\vGridbox@=0pt
  \hbox{\box\hGridbox@\box\vGridbox@}%
  }

 \def\LabelsGlobal{\def\LabGl@b{\global}}
 \def\LabelsLocal{\def\LabGl@b{}}
 \LabelsGlobal 

 \def\SetLabels#1\endSetLabels{%
   \LabGl@b\Labeltoks@={#1()\\}%
   }

 \LabGl@b\Labeltoks@={()\\}

 \def\ShowGrid{\LabGl@b\let\PlaceGrid@\PlaceGrid@@}
 \def\HideGrid{\LabGl@b\let\PlaceGrid@\relax}
 \def\Grids{\ShowGrid\LabGl@b\let\GridSwitch@\ShowGrid}
 \def\noGrids{\HideGrid\LabGl@b\let\GridSwitch@\HideGrid}

 \noGrids

 \def\bAdjust@@{%
     \setbox\auxLabelbox@=\hbox{\raise \dp\auxLabelbox@
            \box\auxLabelbox@}}
 \def\bAdjust@{\let\vAdjust@\bAdjust@@}

 \def\eAdjust@@{\dimen0=-.5\ht\auxLabelbox@
     \advance\dimen0 by .5\dp\auxLabelbox@
     \setbox\auxLabelbox@=
            \hbox{\raise\dimen0\box\auxLabelbox@}}
 \def\eAdjust@{\let\vAdjust@\eAdjust@@}

 \def\tAdjust@@{%
     \setbox\auxLabelbox@=\hbox{\raise-\ht\auxLabelbox@
            \box\auxLabelbox@}}
 \def\tAdjust@{\let\vAdjust@\tAdjust@@}

 \let\vAdjust@\relax

 \def\lAdjust@{\let\hAdjust@\rlap}
 \def\rAdjust@{\let\hAdjust@\llap}

 \let\hAdjust@\relax\let\vAdjust@\relax

 \def\FetchLabel@#1(#2)#3\\{%
     \IN@0#2@@\ifIN@
        \setbox0=\hbox{\ignorespaces#1#3\unskip}%
        \ifdim\wd0>0pt
           \ms@g{}%
           \ms@g{ !!! Bad label(s)? !!!}%
           \message{ #1(#2)#3}%
        \fi
        \def\LabelMole@##1\endFetchLabel@{%
            \IN@0()\\@##1@%
            \ifIN@\def\Temp@{\FetchLabel@##1\endFetchLabel@}%
            \else\def\Temp@{}%
            \fi
            \Temp@
           }%
     \else
       \ignorespaces#1\unskip
       \setbox\auxLabelbox@=%
         \hbox to 0pt{\hss\ignorespaces\hAdjust@
          {\ignorespaces#3\unskip}\hss}%
       \vAdjust@
       \let\hAdjust@\relax\let\vAdjust@\relax
       \AugmentLabelBox@@{#2}%
       \ht\Labelbox@=0pt\dp\Labelbox@=0pt
       \let\LabelMole@\FetchLabel@%
     \fi\LabelMole@}

 \newtoks\XYSep@ 
 \def\SetXYSeparator#1{%
     \IN@0#1@@\ifIN@\XYSep@{*}%
     \else
     \XYSep@{#1}%
     \fi
     }

 \SetXYSeparator*

 \def\AugmentLabelBox@@#1{%
     \IN@0\the\XYSep@ @#1@\ifIN@
       \SPLIT@0\the\XYSep@ @#1@%
       \setbox\Labelbox@=\hbox to 0pt{%
         \unhbox\Labelbox@
         \Shifted@@{\the\Initialtoks@\Wddd@}%
         {\the\Terminaltoks@\Httt@}%
         {\box\auxLabelbox@}}%
     \else
         \ms@g{}%
         \ms@g{ !!! Bad insertion point. !!!}%
         \message{ (#1\ this point was rejected.)}%
     \fi
    }

 \def\FetchOption@#1[#2]#3\endFetchOption@{%
    \def\temp{#1}
    \ifx\temp\empty
       \Edgect@=#2\relax
       \let\EdgGl@b\relax
       \SetEdges@
       \Cleaner@#3%
    \fi}

 \def\Cleaner@#1[@]{\Labeltoks@{#1}}
     
 \def\PlaceLabels@@{\mathsurround=0pt
     \def\Cr@{\\}%
     \let\L\lAdjust@\let\R\rAdjust@
     \let\B\bAdjust@\let\E\eAdjust@\let\T\tAdjust@
     \expandafter\FetchOption@\the\Labeltoks@[@]\endFetchOption@
     \Wddd@=\Wdd@ \Edgescale@\Wddd@ 
     \Httt@=\Htt@ \Edgescale@\Httt@
     \expandafter\FetchLabel@\the\Labeltoks@\endFetchLabel@
     \box\Labelbox@
     }%

 \let \PlaceLabels@\PlaceLabels@@

 \def\AffixLabels#1{\setbox\Coordinatebox@=\hbox{#1}%
      \Wdd@=\wd\Coordinatebox@ \Htt@=\ht\Coordinatebox@
      \advance\Htt@ \dp\Coordinatebox@
      \hbox{\copy\Coordinatebox@\kern-\Wdd@ 
           \Shifted@@{0pt}{-\dp\Coordinatebox@}%
           {\PlaceLabels@\PlaceGrid@}%
           \kern\Wdd@}%
      \GridSwitch@ 
      \LabGl@b\Labeltoks@{()\\}%
      }
 
   \let\wlog\LabelFigwlog@ld   
   \catcode`\@=\LabelFigCatAt  

                                By

              Raymond S\'eroul <A18645@FRCCSC21.BITNET>
                                and 
              Laurent Siebenmann <lcs@topo.math.u-psud.fr>
    
              VERSIONS: July 1991, Oct 1991, Jan 1992, July 1992

INTRODUCTION

      This labelling package is intended for TeX users who
rely on non-TeX sources for for their graphics inserts.  It
provides means for adding TeX labels to such inserts with a
minimum of fuss. 

       For most labels, TeX users have in the past found it
reasonably convenient to rely on non-TeX sources. Typical
occasions when an inescapable need for TeX labels seemed to
arise are

 (a) when the graphics program lacks certain exotic or complex
mathematical symbols

 (b) when the very highest typographical quality is wanted for the
labels

 (c) when labels included with the graphics fail to print, 
 and you cannot figure out why (cf. boxedeps.doc).  The labels
 provided by labelfig.tex are 100

       Since this package first appeared, many users, who in the
past scarcely dreamed of using TeX labels, have come to use
nothing but.  So it is now appropriate to add

Intoxication Warning:  TeX labels may be addictive and expensive. 

     If you have a fast preview you may disagree, and even find
that this package provides an agreeable paste-up environment; see
extra applications at end.

     Note to publishers: It is possible and convenient to ultimately
export the TeX labels produced by labelfig.tex to become an integral
part of the EPS file. This is often desired by a publisher who typically
uses an "upmarket" graphics or page layout program, with which the
staff is skilled in perfecting figures.  See Appendix I for
a recipe.

     The authors are grateful to Patrick Ion of Math Reviews for
helpful comments and encouragement.

BASIC INSTRUCTIONS

    After reading in the macro file using

\input labelfig.tex
preview or proof your figure with a coordinate grid printed on
top, by typing the following:

    \ShowGrid  
    \AffixLabels{<the graphics insertion>}

Here <the graphics insertion> is what you would type to insert
the graphics object alone without the grid.  This must provide
for the space around it. For example <the graphics insertion>
might well be \BoxedEPSF{MyFigure scaled 700} using the
boxedeps.tex macro package (from same source); this provides a
TeX box containing the encapsulated PostScript insert specified by
the file MyFigure. \AffixLabels{...} provides the grid (supposing
\ShowGrid is present) and later, once you have specified labels
using the grid, it will "tack on" the labels.

     The grid is a sort of (usually elongated) checkerboard of
ten rows and ten columns and its (internal) partitions are by
default numbered  .1, ... ,.9  both horizontally (X-coordinate
running left to right) and vertically (Y-coordinate running bottom
to top).  Thus the points enclosed by the grid correspond to the
points of the unit square in the cartesian "X-Y" plane, the lower
left corner corresponding to the origin (0,0).  By extrapolation,
the full page corresponds to a larger rectangle in the plane.

     These coordinates serve to position labels as follows.
Before the \AffixLabels{...} command type label specifications:

  \SetLabels
   (<X-coordinate>*<Y-coordinate>) <first label> \\
   .
   .
   .
   (<X-coordinate>*<Y-coordinate>)  <last label> \\
  \endSetLabels

Each row specifies one label and is terminated by \\.  In each
row, the position indicator comes first; it is written as a
standard cartesian point except that the X- and Y- coordinates
are separated by * rather than a comma because TeX allows a
comma as decimal point. There are no dimension units to specify
as the unit is the grid itself.

     By default, this cartesian point specifies where the middle
of the baseline of the label will be located.  However if you precede
the point by \L [or \R] the left [or right] edge of the baseline will
be located there. Similarly you may also precede the point by \T, \E,
or \B to vertically align the top equator or bottom of the label box
at the specified point.  This gives nine standard positions of
the label with respect to the insertion point --- corresponding to
the eight principle points of the compas and the center

                     \L\T     \T      \R\T

                     \L\E     \E      \R\E

                     \L\B     \B      \R\B

But this neglects the default "baseline" level of TeX,
giving potentially three more positions

                     \L    <no tag>   \R

For text, the baseline level is often the preferred. Its relation to
the others is variable. It will often coincide with the bottom level,
as happens for "X".  But it is often distinct, as for "g", in which
case you have in all 12 distinct positions rather than 9.

     It is convenient to think of this specification of label
position as attaching the label by a thumb-tack to the coordinate
grid. There are up to twelve positions of the thumb-tack on the
label, while the position of the thumb-tack on the coordinate grid is
arbitrary.  Normally, one choses the position of the thumb-tack on
the label to be the one that is the closest to the item being
labeled.  There are good reasons for this "rule of thumb":

   (a)  It facilitates correct positioning at first try.

   (b)  If the scale of the figure must be altered after labels
have been affixed, the labels have a good chance of remaining well
positioned.

   (c)  The visible grid need not extend beyond the "bounding box"
for the figure, because the best preferred position is always
(at least almost) within the bounding box .

The second reason is particularly important. Indeed it often
happens that scale has to be altered after labelling begins, in
order to either provide space for the labels, or to adjust
proportions between the labels and the figure.  (The size of labels
is unaffected by scaling.)

     Here is an artificial but self-contained test which uses
TeX rules to make a graphics object.

TEST

    Do not skip this!

\input labelfig

 \def\FrameIt#1{\hbox{\vrule$\vcenter {\hrule\kern3pt%
             \hbox {\kern3pt #1\kern3pt}%
               \kern3pt\hrule}$\relax\vrule}}

 \def\Caption#1#2{\FrameIt{%
       \vtop {\hsize=#1\relax \parindent=0pt
         \leftskip=0pt \rightskip=0pt plus15pt
         \parfillskip=0pt
         \lineskip=1pt\baselineskip=0pt
         #2}}}

 \def\FirstQuadrant{\hbox to 100pt{\vrule\vbox to 100pt{%
        \hbox to 100pt{\hfil}\vfil\hrule}\hss}}

  \SetLabels
    \R(.5*.2) $\zeta\,\cdot$\\
    (.9*-.10) $\xi$\\
    \R(-.03*.9) $\eta$\\
    \T(.5*.9) \Caption{70pt}{%
          \it The norm of
          $g(\xi+i\eta)$ is indicated on
          contours of this invisible surface.}\\
  \endSetLabels

  \AffixLabels{\FirstQuadrant}

  \end

  Note that the coordinates to use for labels are indicated on the
edges of the grid (when visible) corresponding to the conventional
x- and y- axes of the Cartesian plane. By default the grid is
1-by-1. However, by the command \Edges{100}, you can change this
to 100-by-100 and many users find this alternative most
convenient. Place the command \Edges{...} in your style file (or
header) since its effect is is global. Other possible edge values
are 10 and 1000.

  If you use the command \Edges{...} at all, do so with care.  For
if you accidentally delete an \Edges{...} command your labels will
abruptly be badly misplaced and may logically but mysteriously
generate "dimension too big" errors under TeX and "off page" errors
under your driver.  

  You can dictate the edgescale for an individual figure by giving
the scale in brackets immediately after \SetLabels.  Thus, to
import into an article using say \Edge{100} a figure labelled using
another edgescale, say the original 1-by-1 default, you can use
\SetLabels[1]...\endSetLabels.

GETTING IT DOWN PAT

     Complicated labeling deserves the same respect as
complicated mathematics.  Do not expect it to come out perfect the
first time!  What is needed in either case is a mechanism to
repeatedly typeset troublesome pieces.

     One mechanism is always available.  One does complicated
labelling in a separate "test" file involving just the figure being
labelled;  a texpert will know how to \dump TeX's current state as
a temporary format that restarts rapidly at each retry.  Usually,
one then pastes the completed labelled figure back into the main
TeX file, but, of course, one can also \input it as an auxiliary
file.

     If you do not have a TeXpert at handy, here is a first
approximation to an efficient setup. By deletions reduce a copy
of your article to just a few lines before and after the figure.
Now label the figure, and finally, copy and paste the labelled
figure to the original article. Then copy the next figure to label
into this testbed and repeat. The TeXpert can improve the  speed
at which TeX starts up, by compiling a format specifically for
your article; just one caution: best NOT include in the format
ephemeral details of setup like \Set<mydriver>ArtSpecials (from
boxedeps.tex because this reads  figure dimensions which you may
change during your work session.

     An improved mechanism to repeatedly typeset troublesome
pieces is now available on the Macintosh; it is called LinoTeX;
see the same ftp sources.  It could be set up on many types
of computer.

     Before using labelfig.tex to attach labels to a graphics
object inserted using boxedeps.tex or BoxedArt.tex, make it a
firm rule to carefully adjust the bounding box using the trimming
commands of these packages, and also at least tentatively scale
and position the object. Beware of changing the grid inadvertently
after the labels have been positioned.  For example, correcting
the bounding box of a PostScript graphics object can foul up the
labels by changing the coordinate grid to which the labels are
attached. This is particularly true for the trimming  commands of
boxedeps.tex and BoxedArt.tex. However, as noted already, change
of scale is much less disruptive, and modest adjustments should be
well tolerated.

     Sometimes the labels protrude so far from the bounding box
of a figure that the figure has to be repositioned.  Best do this
by ad hoc spacing, say using \hglue and \vglue; altering the
bounding box would create a vicious circle.

     Remember that you are responsible for preventing labels
from overlapping. You are responsible for all label typography
including size and style. A label is really just about anything
that can be put in a TeX box. Note that spaces at the beginning
and end of labels will normally be suppressed; if you really want
them you must protect them with TeX braces.

     This package temporarily sets the \mathsurround parameter
of TeX to zero  while the labels are being affixed. This is done
because nonzero \mathsurround space would influence the position
of left and right aligned labels; then, when a texpert or printer
modifies mathsurround, diagram labeling might be disastrously
altered. There is a small price to pay involving labels that are
formatted as caption boxes including mathematics: you  may want or
need to specify an explicit mathsurround space within the caption
box; it will not influence anything outside.

     Those hostile to the use of * as separator between
the X and Y coordinates of label insertion points, are free to
impose another using \SetXYSeparator{<the new separator>}.  
Americans may prefer "," to "*" since they never use a 
comma as a decimal point; on the other hand, * may be more visible.

APPENDIX (I)  MERGING labelfig.tex LABELS INTO AN EPSF GRAPHICS OBJECT.

     As promised in the introduction, here is a recipe useful for
publishers. It works at least on Macintosh and at least for vectorized
graphics and Adobe type1 fonts.  (There is surely a similar recipe for
PCs under MSWindows.)

 (a)  Use boxedeps.tex utility to integrate the figure given by the eps
file, "x.eps" say, with a visible frame around it.  See
\ShowDisplacementBoxes command in boxedeps.tex.  To get precise results
automatically it is important to use the \Trim... commands of
boxedeps.tex making the "DisplacementBox" neatly fit the figure.

 (b)  Use the TeX printer driver and LaserWriter (versions >= 8.1.1) to
export to an EPSF the DVI page containing the integrated, labelled
figure. You now have an EPS file  "xx.eps"  that contains too much, and at
the wrong scale, and at wrong position.

 (c)  Convert the EPSF to an Adode Illustrator format EPSF using
the shareware utility called epsConvert by Sam Weiss
1993-- (currently $25).

 (d)  In Illustrator (or a compatible program), group the labels and the
"DisplacementBox"; copy them to the clipboard and paste them into "x.ps".
This step requires that all the label fonts be "visible to the Macintosh.

 (e)  Translate and scale the pasted group consisting of the labels plus
the "DisplacementBox" so as to make the "DisplacementBox" the bounding
box of (labelless) figure represented by "x.eps".  At this point the
labels will be correctly placed on the figure "x.eps".

 (f)  Ungroup and delete the "DisplacementBox".  The result is the
desired single EPS file, "x+.eps" say, It contains the original figure
plus its labels.  

     Using grouping and ungrouping appropriately in "x+.eps", a
publisher's staff can very efficiently improve label positions etc.

APPENDIX II)  SOME EXOTIC APPLICATIONS

     The grid of labelfig.tex is analogous to a light-table in
classical page makeup with wax or latex glue.  In principle, you
can use it to compose any page from its indivisible parts.  This
even has some of the artisanal charm of classical paste-up
provided you have a fast screen preview to make the process
"interactive".

     In practice labelfig.tex is a tool for nonstandard jobs.
Here are a few going beyond the labelling already discussed.

(I)  GRAPHICS INTEGRATION.

     This is accomplished by treating the imported graphics
objects as labels.  The underlying graphics object is then
typically an empty  \vbox to <dimension>{\vfill} in a TeX
\midinsert...\endinsert construction.  A label line
might be of the form

   (.1*.1) \\

The exact form of the special command varies from driver to
driver.  However, in the case of encapsulated PostScript graphics
(EPSF norm), by relying on boxedeps.tex, one can have the
following standard syntax (independant of driver  (see
boxedeps.doc for details.
  
  (.1*.1) \BoxedEPSF{MyFigure scaled <scale in mils>}\\

This may be slow since it requires TeX to read the PostScript
file to read bounding box using many complex macros.  So you
may want to try

  (.1*.1) \EPSFSpecial{MyFigure}{<scale in mils>}\\

which is fast and driver independant, but it squashes the
bounding box, normally to its lower left corner.

     Similarly for graphics of the Macintosh PICT norm ---
using BoxedArt.tex (same sources) in place of boxedeps.tex.

     This approach to integration is to be recommended when
one is assembling a composite graphics object.

 (II)  COMMUTATIVE DIAGRAM ENHANCEMENT

     Commutative diagrams or arrays of mathematical objects
connected by arrows of various sorts are common in mathematics.
The mathematical objects require the use of TeX.  Recently TeX
acquired a good collection of arrows of all slopes --- that of
LamSTeX --- plus pwerful macros to build the diagrams.

     However, even the LamSTeX collection is often
inadequate; it lacks for example double shafted arrows, dotted
arrows and curved arrows. Fortunately it is possible to produce
such arrows on an individual basis using sophisticated graphics
programs such as Illustrator and AldusFreehand (both serving
the EPSF norm) or using Metafont (with its public domain norm).
Since the creation of each new arrow is a work of love, you
probably want to limit the number of arrows by using LamSTeX
for most arrows. The 40K commutative diagram module of LamSTeX
has been adapted to work with AmSTeX and a copy may be posted
with LabelFig and related files. Unfortunately no one has yet
offered a version that works with Plain TeX or LaTeX.

       Suffice it here to say that when the exotic arrow has
been somehow imported into TeX, labelfig.tex treats it as a
label that one affixes to the commutative diagram.  Two other
steps will be treated in separate notes, namely the matter of
extracting the dimension specifications for the arrow and the
construction of the arrow --- for these steps are far from
unique and often depend intimately on your computer environment. 
Notes for the Macintosh-Textures-Illustrator combination are
found in the file ExoticArrows.doc.

 (III) NESTING 

Ingenuity pays off in exploiting labelfig.tex. One can
mix graphics and typography quite freely.  labelfig.tex is good
for freeform or overlapping arrangements, while boxedeps.tex (or
BoxedArt.tex) is best for regimented non-overlapping
arrangements --- and the two can be combined.

     The default behavior of labelfig.tex is not ideal 
for nesting objects, because to prevent trouble for beginners
the register for labels is globally cleared when \AffixLabels
concludes.  But there are switches available

      \LabelsGlobal      \LabelsLocal

which change this.  To understand this, extend the above test 
by something like:

 \LabelsLocal

 \SetLabels
    (.5*.5) AAA\\
 \endSetLabels

 {
 \SetLabels
    (.5*.5) ZZZ\\
 \endSetLabels
   \AffixLabels{\FirstQuadrant}
 }

   \AffixLabels{\FirstQuadrant}

     There are however potential pitfalls.  Neither
labelfig.tex nor boxedeps.tex has been tested under extreme
conditions. Problems may occur if their procedures are
indiscriminately nested. For boxedeps.tex (not labelfig.tex)
there is a precise cause for worry, namely many of its
variables are "global", which means that TeX braces will not
provide the protection one might expect.

COMMAND SUMMARY FOR labelfig.tex

  Here [...] means optional (one or zero)
       [...]* means any number of such constructs

  \SetLabels
    [[<P>](<X><Sep><Y>) <label> \\]*
  \endSetLabels
  \ShowGrid  
  \AffixLabels{<the figure>}

   --- <P> is tack position, one of eleven or empty
              order irrelevant

                   \L\T      \T      \R\T

                   \L\E      \E      \R\E

                     \L               \R

                   \L\B      \B      \R\B

   --- (<X><Sep><Y>) insertion point;
  <Sep> is separator, = * by default;
  \SetXYSeparator{<Sep>} changes it.
   <X> and <Y> are real numbers

  --- <label> a label to attach 

  --- <the figure> the figure to label 

  \GlobalLabels (default)     
  \LocalLabels  setting for nested constructs.

 \Grids makes ALL grids appear; \HideGrid then makes just next disappear.
 \noGrids returns to default.  The commands are always global.

 \GridLineWidth{<dimension>} adjusts width of grid lines. Default is very
small, to give "hairline" effect. If your grid lines are missing try
setting \GridLineWidth{1pt}.

 \Edges#1 globally changes the edge size of all grids to the numerical 
value #1, which must be 1, 10, 100, or 1000.  The default is 1.

VERSION HISTORY.
 --- Jan 1993: \Edges#1 and [??] option after \SetLabels
 --- July 1992: \Grids, \noGrids, \HideGrid;
       Gridlines become hairlines; \GridLineWidth{<dimension>}.
 --- Oct 1991, Jan 1992: \SetXYSeparator{<Sep>},  \LabelsGlobal,
       \LabelsLocal.
 --- July 1991: first release

Address for bugs and other feedback:

        Raymond S\'eroul
        IREM and Lab. de Typographie Informatise
        Univ. Rene Descartes
        Strasbourg

    Tel 33-88-41-63-45
    Email:  A18645@FRCCSC21.BITNET

        Laurent Siebenmann
        Mathematique, Bat. 425,
        Univ de Paris-Sud,
        91405-Orsay,
        France

    Tel 33-1-6941-7949; 
    Email: lcs@topo.math.u-psud.fr

%% file: paper_v8.bbl
\begin{thebibliography}{}


\bibitem{3gpp:NOMA} G. T. 36.859, ``Study on downlink multiuser superposition transmission,"
{\it 3GPP}, 2015.

\bibitem{Saito:13VTC} Y. Saito, Y. Kishiyama, A. Benjebbour, T. Nakamura, A. Li, and K. Higuchi, ``Non-orthogonal multiple access (NOMA) for cellular future radio access," in {\it Proc. IEEE VTC Spring}, pp. 1-5, Jun. 2013.

\bibitem{Ding:14SPL} Z. Ding, Z. Yang, P. Fan, and H. V. Poor, ``On the performance of non-orthogonal multiple access in 5G systems with randomly deployed users," {\it IEEE Trans. Signal Process.}, vol. 21, pp. 1501-1505, Jul. 2014.

\bibitem{Timotheou:15SPL} S. Timotheou and I. Krikidis, ``Fairness for non-orthogonal multiple
access in 5G systems," {\it IEEE Signal Process. Lett}, vol. 22, pp. 1647-1651, Mar. 2015.

\bibitem{Liu15:PIMRC} F. Liu, P. M{\"a}h{\"o}nen, and M. Petrova, ``Proportional fairness-based user
pairing and power allocation for non-orthogonal multiple access," in {\it Proc. IEEE PIMRC} pp. 1127-1131, Aug. 2015.

\bibitem{So&Sung:15ComLet}  J. So and Y. Sung, ``Improving non-orthogonal multiple access by
forming relaying broadcast channels," {\it IEEE Communications Letters}, vol. 20, pp. 1816-1819, Sep. 2016.

\bibitem{Kim13:MILCOM} B. Kim, S. Lim, H. Kim, S. Suh, J. Kwun, S. Choi, C. Lee, S. Lee, and D. Hong, ``Non-orthogonal multiple access in a downlink multiuser beamforming system," in {\it Proc. MILCOM}, pp. 1278-1283, 2013.

\bibitem{Lan14:ICSPCS} Y. Lan, A. Benjebboiu, X. Chen, A. Li, and H. Jiang, ``Considerations on downlink non-orthogonal multiple access (NOMA) combined with closed-loop SU-MIMO," in {\it IEEE ICSPCS, 2014 8th International Conference on}, pp. 1-5, 2014.

\bibitem{Liu15:ICCW} S. Liu, C. Zhang, and G. Lyu,, ``User selection and power schedule for downlink non-orthogonal multiple access (NOMA) system," in {\it Proc. IEEE ICCW}, pp. 2561-2565, 2015.

\bibitem{Hanif16:TSP} M. F. Hanif, Z. Ding, T. Ratnarajah, and G. K. Karagiannidis, ``A minorization-maximization method for optimizing sum rate in the downlink of non-orthogonal multiple access systems," {\it IEEE Trans. Signal Process.}, vol. 64, pp. 76-88, Jan. 2016.

\bibitem{Ding&Adachi&Poor:16WC} Z. Ding, F. Adachi, and H. V. Poor,, ``The application of MIMO to nonorthogonal multiple access," {\it IEEE Trans. Wireless Commun.}, vol. 15, pp. 537-552, Jan. 2016.

\bibitem{Sayed17:WD} A. Sayed-Ahmed and M. Elsabrouty, ``User selection and power allocation for guaranteed SIC detection in downlink beamforming nonorthogonal multiple access," in {\it Proc. IEEE Wireless Days}, pp. 188-193, 2017.

\bibitem{Seo&Sung:18SP} J. Seo and Y. Sung, ``Beam design and user scheduling for nonorthogonal multiple access with multiple antennas based on Pareto optimality," {\it IEEE Trans. Signal Process.}, vol. 66, pp. 2876-2891, Jun. 2018.

\bibitem{Chen16:Access} Z. Chen, Z. Ding, and X. Dai, ``Beamforming for combating inter-cluster and intra-cluster interference in hybrid NOMA systems," {\it IEEE Access}, vol. 4, pp. 4452-4463, Aug. 2016.

\bibitem{Chen16:TSP} Z. Chen, Z. Ding, X. Dai, and G. K. Karagiannidis, ``On the application of quasi-degradation to MISO-NOMA downlink," {\it IEEE Trans. Signal Process.}, vol. 64, pp. 6174-6189, Aug. 2016.

\bibitem{Yoo&Goldsmith:06Jsel} T. Yoo and A. Goldsmith, ``On the optimality of multiantenna broadcast scheduling using zero-forcing beamforming," {\it IEEE J. Sel. Areas in Commun.}, vol. 24, pp. 528-541, Mar. 2006.

\bibitem{Choi15COM} J. Choi, ``Minimum power multicast beamforming with superposition coding for multiresolution broadcast and application to NOMA systems," {\it IEEE Trans. Commun.}, vol. 63, pp. 791-800, Mar. 2015.

\bibitem{Harashima&Miyakawa:72TCOM} H. Harashima and H. Miyakawa, ``Matched-transmission techinique for
channels with intersymbol interference," {\it IEEE Trans. Commun.}, vol. 20, pp. 774-780, Aug. 1972.

\bibitem{Tomlinson:71EL} M. Tomlinson, ``New automatic equaliser employing modulo arithmetic," {\it Electron. Lett.}, vol. 7, pp. 138-139, Mar. 1971.

\bibitem{Windpassinger&Fischer&Vencel&Huber:04WC} C. Windpassinger, R. F. H. Fischer, T. Vencel, and J. B. Huber, ``Precoding in multiantenna and multiuser communications," {\it IEEE Trans. Wireless Commun.}, vol. 3, pp. 1305-1316, Jul. 2004.

\bibitem{Tse&Viswanath:book} D. Tse and P. Viswanath, {\it Fundamentals of Wireless Communications.} Cambridge, U.K.: Cambridge University Press, 2005.

\bibitem{Scutari&Facchinei&Lampariello:17SP} G. Scutari, F. Facchinei, and L. Lampariello, ``Parallel and distributed methods for constrained nonconvex optimization - Part I: Theory," {\it IEEE Trans. Signal Process.}, vol. 65, pp. 1929-1944, Apr. 2017.

\bibitem{Li&Chang&Lin&Chi:13SP} W.-C. Li, T.-H. Chang, C. Lin, and C.-Y. Chi, ``Coordinated beamforming for multiuser MISO interference channel under rate outage constraints," {\it IEEE Trans. Signal Process.}, vol. 61, pp. 1087-1103, Mar. 2013.


\end{thebibliography}
